\documentclass[twocolumn,showpacs,preprintnumbers,amsmath,amssymb,pre]{revtex4-1}

\usepackage{graphicx}
\usepackage{dcolumn}
\usepackage{overpic}
\usepackage{float} 
\usepackage{bbm}

\newcommand{\myD}{ \mathbf{d}}
\newcommand{\myC}{\mathbf{c}}

\begin{document}

\title{Bosonic Josephson effect in the Fano-Anderson model} 

\author{ G. Engelhardt    }
\email{georg@itp.tu-berlin.de}

\author{G. Schaller}

\author{ T. Brandes}

\affiliation{%
Institut f\"ur Theoretische Physik, Technische Universit\"at Berlin, Hardenbergstr. 36, 10623 Berlin, Germany }

\begin{abstract} 
	We investigate the coherent dynamics of  a non-interacting Bose-Einstein condensate in a system consisting of two bosonic reservoirs coupled via a spatially localized mode.
	We describe this system by a two-terminal Fano-Anderson model and investigate analytically the time evolution  of observables such as the bosonic Josephson current. In doing so, we find that the Josephson current  sensitively depends on  the on-site energy of the localized mode. This facilitates to use this setup as a transistor for a Bose-Einstein condensate.
	 We identify two regimes. In one  regime, the system exhibits  well-behaved long-time dynamics with a slowly oscillating and undamped Josephson current. In a second regime,  the Josephson current is a superposition of an extremely weakly damped slow oscillation and an undamped fast oscillation. Our results are confirmed by finite-size simulations.
\end{abstract}


\maketitle
%%%%%%%===========================================================================================================

 %%%
 %%%
 \begin{figure*}[t]
   \centering
   \includegraphics[width=1\linewidth]{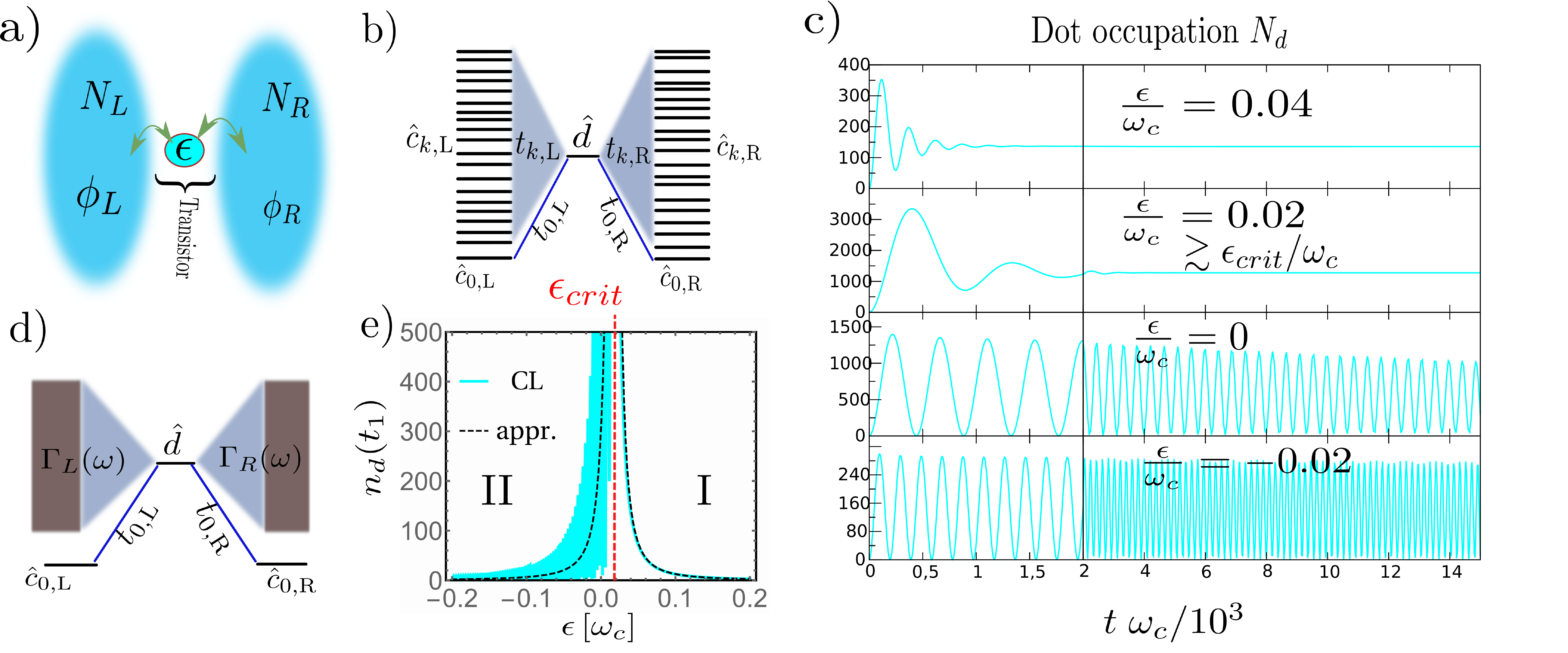}
   \caption{(Color online) 
   \textbf{(a)} Sketch of the  system. Two reservoirs are coupled via a strongly confined well.
    \textbf{(b)} The system is described by the Fano-Anderson model in Eq.~\eqref{eq:Hamiltonian}. 
    \textbf{(c)} Dot occupation $N_d$ as  function of time for different gate potentials $\epsilon$. For $\epsilon> \epsilon_{crit}\approx 0.016 \omega_c$, the occupation is constant, while it is  oscillating for $\epsilon<\epsilon_{crit}$ in the long-time limit.
   \textbf{(d)} In our calculations, we formally exclude the coupling of the lowest energy modes $\myC_{0,\alpha}$ in the reservoirs $\alpha=L,R$ from the continuum limit (CL) to take into account the dynamics of the BEC.
    \textbf{(e)} Dot occupation $N_d$ as  function of $\epsilon$ for a fixed time $t_1 \omega_c= 2\cdot 10^3$. The occupation is extremely high close to the transition at $\epsilon= \epsilon_{crit}$.
    The black line depicts the analytic time-averaged dot occupation in Eq.~\eqref{eq:dotOccupation}.
     The overall parameters are $n_{L}=n_{R}=10^4$, $\Delta\phi=-\pi/2$, $\eta=0.5$, $\gamma_\alpha/\omega_c = 0.028$, and $t_{0,\alpha}/\omega_c= 0.0021$ and $\omega_{0,\alpha}=0$.  For more details about the choice of  $t_{0,\alpha}$ see appendix~\ref{sec:AppFiniteSizeSimulation}. 
   }
   \label{fig:overview}   
   \end{figure*}
 %%%
 %%%

 \section{Introduction}
 
 The experimental control of cold-atomic quantum gases has proceeded to a high level in the recent years. In particular, transport experiments in  two-terminal setups exhibit interesting effects as, e.g., conductance quantization or the creation of a heat engine \cite{Krinner2015,Brantut2012,Brantut2013}. Moreover, the control  of  superfluids gives rise to new transport regimes: in contrast to   particle transport driven by a difference of the chemical potentials in the reservoirs or by a temperature gradient, the dynamics of a superfluid is determined by the phase of its matter wave \cite{Levy2007,Jendrzejewski2014,Albiez2005}. The control of these kinds of systems could pave the way to establish so-called atomtronic circuits \cite{Ryu2015,Stickney2007,Pepino2009}. 
 
 Moreover, theoretical investigations of bosonic transport predict interesting effects as, e.g., a quantization of the current, superfluid Helmholtz oscillations or current against the chemical potential gradient~\cite{Papoular2014,Papoular2015,Nietner2014,Schaller2014,Gallego-Marcos2014,Kordas2015,Haag2015}. However, all of them rely to some extend on phenomenological assumptions to describe the many-particle dynamics, as interactions destroy integrability.
 
 Motivated by these experimental and  theoretical achievements, we study here the dynamics of a system which consists of two bosonic reservoirs which may at low temperature contain Bose-Einstein condensates (BECs).
  These reservoirs are linked by an additional strongly confined potential well. The situation is sketched in Fig.~\ref{fig:overview}(a). 
  For weak couplings and low temperature, the dynamics is mainly governed by a bosonic Josephson effect~\cite{Josephson1962,Likharev1979,Gati2007}. Thereby, the Josephson current between two directly coupled BECs depends on the phase difference of the condensates. Here we investigate, how  the indirect coupling via the additional potential well influences the dynamics. 
   Furthermore, a special focus of this article is the influence of the excited reservoir modes on the dynamics of the BEC. Based on our model and our methods, we are able to analytically investigate the effects of particle loss and damping due to these  excited reservoir modes.
   
  In contrast to the theoretical investigations in Refs.~\cite{Papoular2014,Papoular2015,Nietner2014,Schaller2014,Gallego-Marcos2014,Kordas2015,Haag2015}, we maintain the integrability in our investigation by neglecting the inter-particle interactions. For instance, in Rubidium condensates, the interaction is  rather small and can be additionally adjusted using Feshbach resonances~\cite{Widera2004,Erhard2004}.  The absence of interactions allows us to analytically solve the dynamics of the full system. Here, the only additional assumption enters  by presupposing  the excited reservoir states to be thermally occupied initially. We describe this system as a Fano-Anderson model, which allows for analytical calculations.
  Our aim is to understand the effects in the non-interacting model in detail, which can then provide a starting point for future investigations of the behavior in the  presence of interactions. In particular, we derive an effective non-hermitian Hamiltonian describing exactly the dynamics of the condensate in the long-time limit. Our methods could be thus employed to microscopically  study effects which appear in non-hermitian Hamiltonians as, e.g., the so-called exceptional points, where two eigenvalues and  their eigenstates merge as a function of system parameters~\cite{Berry2004,Heiss2012,Gao2015,Doppler2016}.

 An indispensable device in integrated electronic circuits is a transistor. By adjusting a gate potential, one can control the current from a source to a drain region with high accuracy.  
We investigate the dynamics of the BEC in a bosonic Fano-Anderson model in order to test if  a bosonic system can be applied as a transistor-like device controlling the Josephson current. 
 The confined potential well is assumed to have a large trapping frequency so that it is justified to consider it as single mode with energy $\epsilon$. For this reason, we call the confined potential well  a `bosonic quantum dot'. In the following, due to the close relation to a common transitor, we  denote $\epsilon$ with the gate potential. We demonstrate that   the Josephson current from the left to the right reservoir sensitively depends on $\epsilon$. 
  Furthermore, our approach reveals two regimes in the dynamics induced by the excited reservoir modes. Depending on the gate potential, there is a regime with a  constant dot occupation   in the long-time limit, and a regime where the dot occupation exhibits fast oscillations which persists for very long times. At the transition, the dot occupation is exceedingly high. The Josephson current exhibits a similar behavior. The regimes appear as the energy of the reservoir modes is bounded at energy $\omega=0$. As the energy of the eigenstates of the system generating the dynamics is below or above that boundary, their dynamics is subjected to damping or not.
  
  The structure of the article is as follows. In Sec.~\ref{sec:JosephsonEffect}, we give a general introduction to the Josephson effect in superconducting and bosonic systems. In Sec.~\ref{sec:Model}, we explain the bosonic Fano-Anderson model, for which we specify the Josephson current in Sec.~\ref{sec:JosephsonCurrentFAmodel}. In Sec.~\ref{sec:EoMLaplaceSpace}, we apply the so-called equations-of-motion method to calculate the dynamics, which is discussed in Sec.~\ref{sec:DynamicalRegimes}. Section~\ref{sec:LongTimeLimit} focuses on the dynamics for long times.
  In Secs.~\ref{sec:branchCutAppr}-\ref{sec:EffectiveHamiltonian},  we explain how to efficiently calculate the time evolution and show how to derive an effective Hamiltonian resembling exactly the dynamics in the long-time limit. We provide an exact  expression for the Josephson current in Sec.~\ref{sec:RelationObsRoots}. In Sec.~\ref{sec:LowFrequencyCurrent}, we discuss the low-frequency current which is mainly responsible for the particle transport. The  appendix provides details about the calculations.

 \section{The system and basics }

  \subsection{The Josephson effect}

  \label{sec:JosephsonEffect}
  
  In a conventional superconductor, the Cooper-pairs  form a condensate whose macroscopic order parameter is described by a phase $\phi$. Two superconducting regions connected by a small normal-conducting island constitute a Josephson junction. The coherent Josephson current through this junction depends on the phase difference $\Delta\phi= \phi_R-\phi_L$
  of the condensate phases in the superconducting leads, namely
  \begin{equation}
  I_J(t)= I_c \sin \Delta \phi(t),
  \label{eq:JosephsonCurrent}
  \end{equation}
  where $I_c$ is called the critical current~\cite{Josephson1962,Likharev1979}. In the ac-Josephson effect, the two leads are subjected to a chemical potential bias $\Delta \mu$ which
  gives rise to a time evolution of the phase difference
  \begin{equation}
  \frac{d}{dt}\Delta \phi (t) = \frac{2e}{\hbar} \Delta \mu,
  \label{eq:DeltaPhiTE}
  \end{equation}
  where $2e$ is the charge of a Cooper pair.
  For a constant bias $\Delta\mu$, this results in a  sine-modulated Josephson current with the so-called characteristic frequency of the junction $\Omega_J =\frac{2e}{\hbar} \Delta \mu$. 
  
  An analogue effect appears also in a BEC, whose macroscopic order parameter is the complex-valued wave function. The simplest model with a bosonic Josephson current  consists of two coupled bosonic modes
  \begin{equation}
 	  H = t_0 \left( \mathbf a^\dagger \mathbf b  +  \mathbf b^\dagger \mathbf a \right) 
 	  \label{eq:2ModeHam},
  \end{equation}
  where $t_0$ denotes the tunneling coupling \cite{Gati2007}.
  The bosonic Josephson current related to the operator $\mathbf I \equiv i \left[H, \mathbf b^\dagger \mathbf b\right]$  reads
  \begin{equation}
  I_J (t)=  t_0\; 2\;  \text{Im}\;\left<  \mathbf b^\dagger \mathbf a \right>_t .
  \label{eq:Josephson2mode}
  \end{equation}
   For the initial state $\left|\psi_0 \right>$ at time $t=0$ we assume a BEC described by a product of coherent states  $\left|\psi_0 \right>=\left|a \right> \otimes\left| b \right>$  with $\mathbf a\left|a \right> = \sqrt{n_a}e^{-i\phi_a}   \left|a \right>   $ and $\mathbf b\left|b \right> = \sqrt{n_b}e^{-i\phi_b}   \left|b \right>   $. Thereby, $n_\alpha$ denotes the initial occupation of mode $\alpha=a,b$ and $\phi_\alpha$ its phase.  Solving  the equations of motion we  find
   \begin{align}
   	I_J(t)  =&- \frac 12 \Omega_J   \; \sin(\Omega_J \;t )  \left( n_b - n_a\right)  \nonumber\\
 	  		&+ \Omega_J\;\cos(\Omega_J t )  \sin \Delta \phi_0 \sqrt{n_a n_b},
   		\label{eq:Current2mode}
   \end{align}
 	  where $\Delta \phi_0 = \phi_b - \phi_a$ and  $ \Omega_J= 2 t_0$ is the characteristic frequency. 
    In contrast to \eqref{eq:DeltaPhiTE},  we do not assume a difference in the chemical potentials in the bosonic Hamiltonian as finite energy terms  $\omega_a \mathbf a^\dagger \mathbf a$ and $\omega_b \mathbf b^\dagger \mathbf b$  are not present in Eq.~\eqref{eq:2ModeHam}.   Obviously, the Josephson current exhibits an oscillating behavior with the characteristic  frequency $ \Omega_J= 2 t_0$. This is exactly the energy difference of the two eigenenergies $\epsilon_\pm = \pm t_0$ of the Hamiltonian~\eqref{eq:2ModeHam}. The superposition of the corresponding eigenmodes thus drives the coherent dynamics of the Josephson current. The same effect also generates the BEC dynamics in the Fano-Anderson model considered in this article, although here the coherent dynamics is subjected to incoherent  particle loss due to the coupling to the excited reservoir modes. 
     In Sec.~\ref{sec:LowFrequencyCurrent}, we derive a  relation similar to \eqref{eq:Current2mode} for the low-frequency Josephson current in the Fano-Anderson model, c.f. Eq.~\eqref{eq:CurrentApprox}.
    We find, that the excited reservoir modes effectively  create an additional imaginary part to the energies and to the characteristic frequency $\Omega_J$.

 \subsection{Our model}
 
 \label{sec:Model}
 
We theoretically model the transport system depicted in Fig.~\ref{fig:overview}(a) as a bosonic two-terminal Fano-Anderson model to investigate the transport properties of a BEC which, to our knowledge, has not been done before. The model is
 sketched in  Fig.~\ref{fig:overview}(b).
The Hamiltonian reads
\begin{align}
	\mathbf H = \epsilon \myD^\dagger \myD + \sum_{\alpha=L,R}\sum_{k=0}^{k_{max}}  &\left[ \omega_{k\alpha} \myC_{k \alpha}^\dagger \myC_{k\alpha} \right.  \nonumber\\ 
	&+ \left. \left(t_{k\alpha}\myD^\dagger \myC_{k\alpha}+\text{h.c.}\right)\right],
	\label{eq:Hamiltonian}
\end{align}
where  $\omega_{k \alpha}$ denote  the energy  of the  bosonic reservoir modes $\myC_{k\alpha}$ which are labeled by $\alpha=L,R$ denoting the reservoirs and $k=0, ...,k_{max}$ denoting their internal  states. The parameters $t_{k \alpha}$ describe the coupling of the dot $\myD$ to each reservoir mode. Without loss of generality, we assume real-valued $t_{k \alpha}$ throughout the article. Complex tunneling elements can be rendered real by an appropriate gauge transformation.
 This and related models have been frequently used to study transport in various contexts~\cite{Gurvitz1991,Gurvitz2015,Meir1992,Meir1993,Bruderer2012,Schaller2009}. 
 For a bosonic Fano-Anderson model at temperatures above the condensation temperature, one can show that the stationary particle current is given by~\cite{Topp2015}
\begin{equation}
	I_{R} = \int_0^\infty \mathcal G(\omega) \left[ n_L(\omega)- n_R(\omega) \right] d\omega,
	\label{eq:currentM}
\end{equation}
 where $n_\alpha(\omega)=1/ \left[ e^{\beta_\alpha (\omega -\mu_\alpha)}-1 \right]$ is the Bose distribution and describes the occupation of the left and right reservoir modes. It depends on the chemical potentials $\mu_\alpha<0$ and  the temperatures $T_\alpha= 1/ \beta_\alpha$. The transmission  $\mathcal G(\omega)$ is a system property and does not depend on either temperature or chemical potential. Thus, the current through the system is generated by a difference of the chemical potentials or temperatures in the reservoirs.
 In contrast, we are interested in the  coherent contributions to the  particle transfer which --- just as in the bosonic two-mode system of Sec.~\ref{sec:JosephsonEffect} --- can be present even in the case of vanishing temperature and chemical potential difference.

   \section{Dynamcis of the Bose-Einstein condensate}
   
 \subsection{ Josephson current in the Fano-Anderson model}
 
 \label{sec:JosephsonCurrentFAmodel}
 
 We define the current operator via the time evolution of the particle-number operator of the right reservoir $\mathbf N_R = \sum_k \myC_{k,R}^\dagger  \myC_{k,R}$, thus,
 \begin{align}
 \mathbf I_R &\equiv i \left[\mathbf H, \mathbf N_R \right] = - i \sum_{k} t_{k,R} \myC_{k,R}^\dagger \myD + \text{h.c.} \nonumber \\
			 &\equiv \mathbf I_{R,ex} + \mathbf I_{R,J}.
 \end{align}
 The current can be split into two parts $\mathbf I_{R,ex}$ and  $\mathbf I_{R,J}$. They are related to the initial condition which we specify in the following.

 The density matrix at time $t=0$ is given by a product of
 the density matrices describing  each reservoir separately. The excited states are assumed to be initially thermally occupied. Moreover, we allow for a  condensate in each reservoir $\alpha$ so that the lowest energy modes $\omega_{0,\alpha}$ are macroscopically occupied. This effect requires a finite energy gap between the lowest energy mode and the excited modes. If the temperature of the reservoir is lower than a critical temperature which depends on the particle density, the Bose-Einstein condensation takes place.
 
 Each condensate in the modes $\myC_{0,\alpha}$ is characterized by a macroscopic occupation $n_\alpha$ and a phase $\phi_\alpha$. We  describe it by a coherent state $\left| \alpha_0 \right>$. For these reasons, the initial density matrix reads
 \begin{align}
	 \rho &= \rho_L \otimes\rho_R \otimes \rho_d \label{eq:InitalCondition1}, \\
	 \rho_\alpha &= \frac 1{ Z_\alpha} \exp\left[-\beta_\alpha \sum_{k\neq 0}^{k_{max}} \omega_{k,\alpha}\myC_{k,\alpha}^\dagger \myC_{k,\alpha}\right] \otimes \left| \alpha_0\right>\left< \alpha_0 \right|, \\
	 \myC_{0\alpha} \left| \alpha_0\right> &=  \alpha_0 \left| \alpha_0\right> \quad  \text{with} \quad\left< \alpha_0\right.\left| \alpha_0\right>=1,\\
	 \rho_d &=\left|0 \right> \left<0\right|,
 \end{align}
 where $\alpha_0 =\sqrt{ n_\alpha} e^{-i \phi_\alpha} $, $\beta_\alpha$ denotes the inverse temperature, and $\left|0 \right>$ is the vacuum state of the dot. The excited modes are described by a density matrix of a grand-canonical ensemble. As we consider a BEC, we assume  vanishing chemical potentials $\mu_\alpha\rightarrow 0$. The partition function $Z_\alpha$ warrants  the normalization $\text{tr} \rho_\alpha=1$. As the experiments in Refs.~\cite{Levy2007,Shin2004} demonstrate, the initial ground-state occupations $n_\alpha$ and the phases $\phi_\alpha$ can be controlled with high accuracy.

Accordingly, we split the current into two contributions.
 The first one is the current operator from and to the excited states $k>0$, thus
  \begin{equation}
  \mathbf I_{R,ex} = - i \sum_{k\neq 0} t_{k,R} \myC_{k,R}^\dagger \myD + \text{h.c.} \quad .
  \end{equation}
  Particles which are thermally excited at $t=0$ generate a current given by Eq.~\eqref{eq:currentM} in the long-time limit which is the main part of $\mathbf I_{R,ex} $. Furthermore, particles which are initially in the condensate do not necessarily stay there. During the time evolution they can jump into the dot and then into an excited mode. Thus, also a fraction of the condensate particles  can participate in $\mathbf I_{R,ex} $.

 In the presence of a condensate we identify the current from and to the reservoir ground states $k=0$ as the Josephson current which is coherent. The corresponding current operator reads
   \begin{equation}
   \mathbf I_{R,J} = - i  t_{0,R} \myC_{0,R}^\dagger \myD + \text{h.c.}\quad .
   \end{equation}
 In this article, we are  interested in the latter  contribution. We therefore assume the zero-temperature limit $T_\alpha\rightarrow0$, or equivalently $\beta_\alpha\rightarrow \infty$, where all particles are initially condensed within the lowest energy
 modes $\myC_{0\alpha}$.

 \subsection{Equations of motion in Laplace space}
  \label{sec:EoMLaplaceSpace}
 
    %%%
    %%%
    \begin{figure*}[t]
      \centering
      \includegraphics[width=1\linewidth]{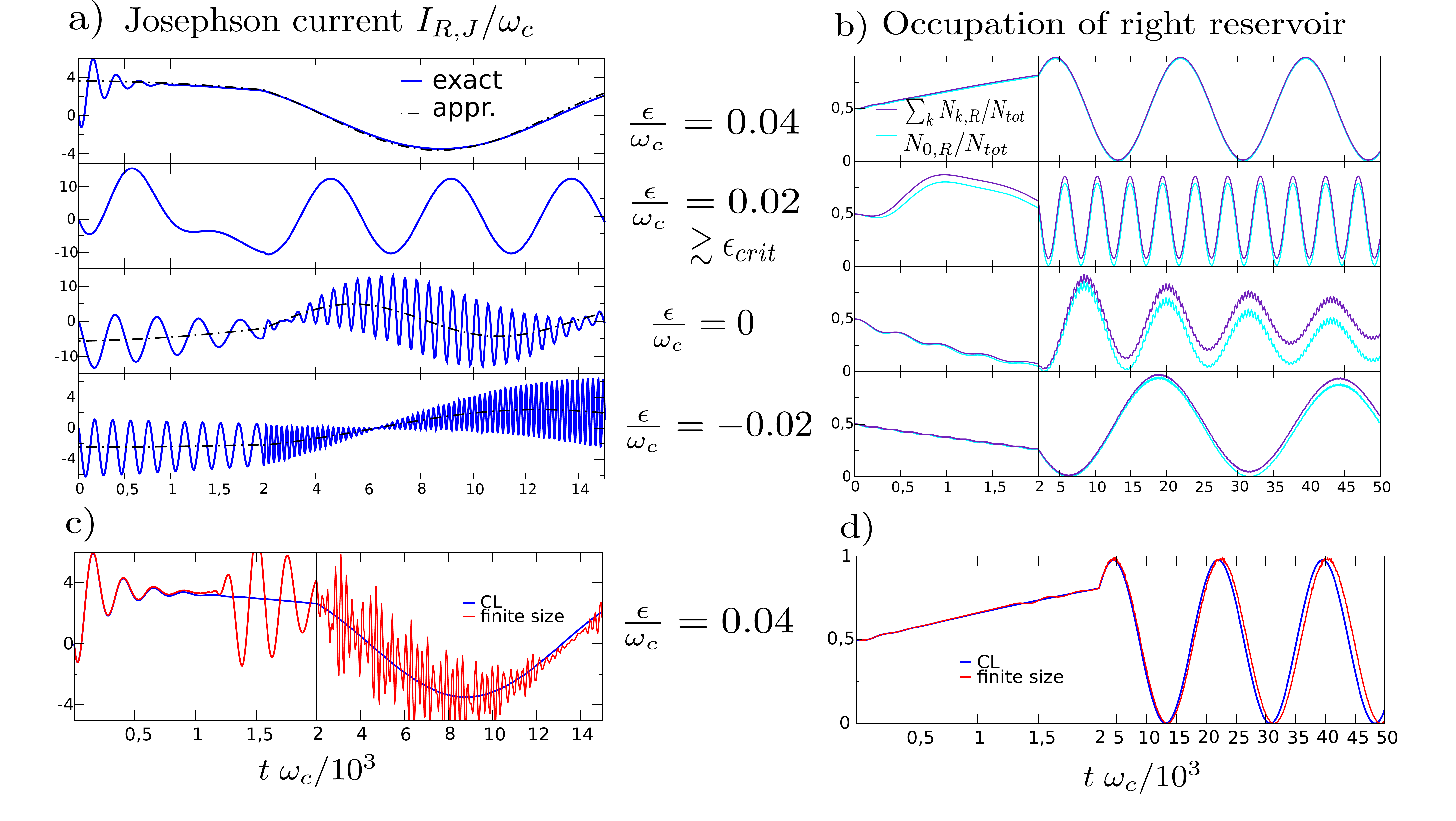}
      \caption{(Color online) 
       \textbf{(a)} Josephson current as a function of time. The parameters are the same as in Fig.~\ref{fig:overview}(c). The current shows a similar behavior as the dot occupation $N_d(t)$. The dash-dotted line depicts an approximation of the time-averaged current given in Eq.~\eqref{eq:CurrentApprox}. For $\epsilon\approx \epsilon_{crit} $ the approximation completely fails (not shown). The oscillation frequencies are given by the imaginary parts of the roots of $\mathcal{\tilde F}(z)$ which are depicted in Fig.~\ref{fig:roots}(e) and are approximately given by Eq.~\eqref{eq:rootApprox}.
       \textbf{(b)} Total occupation $\sum_{k=0}^{k_{max}} N_{k,R}$ and the ground-state occupation $N_{0,R}$ of the right reservoir. Both observables are scaled by the total particle number in the system $N_{tot}=n_L+n_R$. The fast oscillations of the current are only visible in the reservoir occupation close to the transition at  $\epsilon\lesssim \epsilon_{crit}$.
       \textbf{(c)} Finite-size simulation with $k_{max}=100$.  For $t \omega_c\lesssim  1\cdot 10^3$ the finite-size time evolution exactly agrees with the CL. The fluctuations for
              $t\gtrsim \omega_c \cdot 10^3 $ are due to the finite energy gaps between the reservoir modes near the ground state. \textbf{(d)} The fluctuation are  hardly visible in the ground-state occupation $N_{0,R}(t)$.
      }
      \label{fig:Current}
      \end{figure*}
    %%%
    %%%

 %%%
 Following Ref.~\cite{Topp2015}, we  construct the Heisenberg equations of motion $\frac d{dt} \mathbf o = i \left[\mathbf H, \mathbf o \right] $ for $ \mathbf o =   \myD,\myC_{k\alpha} $ and apply a Laplace transformation $\hat {\mathbf o}(z)\equiv \int_0^\infty  e^{-z t}\mathbf o (t)dt$. In Laplace space, the equations of motions can be easily solved and we obtain
 \begin{align}
  \hat {\mathbf{d}}(z) &= \frac{d}{\mathcal F(z) } - i  \sum_{\alpha=L,R} \sum_{k=0}^{k_{max}} \frac{
 t_{k, \alpha}  c_{k, \alpha} }{ \left(  z + i \omega_{k, \alpha} \right)\mathcal F(z) }   \label{eq:SolutionLaplaceD}   ,\\ 
    \hat { \myC}_{k,\alpha}(z)&= \frac{c_{k,\alpha} }{z + i \omega_{k,
 \alpha}}-  \frac{i t_{k, \alpha} d   }{ \left(  z + i \omega_{k, \alpha} \right)
 \mathcal F(z) }  \nonumber \\
                                    &- \sum_{\alpha'=L,R}  \sum_{k'=0}^{k_{max}} 
 \frac{t_{k,\alpha}t_{k',\alpha'}}{\left(  z + i \omega_{k, \alpha} \right) \left(
  z + i \omega_{k', \alpha'} \right)  }  
                                      \frac{c_{k',\alpha'}}{ \mathcal F(z)},
                                      \label{eq:SolutionLaplaceC}
  \end{align}
 %%%
 where $d=\mathbf{d}(t=0)$, $c_{k,\alpha}=\myC_{k\alpha}(t=0)$, and 
 \begin{equation}
 \mathcal F(z)=  z + i \epsilon    + \sum_{\alpha=L,R}\sum_{k= 0}^{k_{max}} \frac{
 t_{k,\alpha}^2}{z + i \omega_{k, \alpha}} .
 \label{eq:characteristicFunction}
 \end{equation}
 The roots of $\mathcal F(z)$ are related to the energies of the Hamiltonian \eqref{eq:Hamiltonian} by
 $\epsilon_i = i z_i$. The time evolution can be obtained by an inverse Laplace transformation 
 \begin{equation}
 	\mathbf o(t) =\frac 1{2\pi i} \int_{\delta-i \infty}^{ \delta+i \infty} e^{ z t}\hat{ \mathbf o }(z) dz ,
 	\label{eq:InverseLaplaceTrafo}
 \end{equation}
 where $\delta>0$ has to be chosen so that the integration contour is completely within the region of convergence of $\hat{ \mathbf o }(z)$.
  In order to perform analytical calculations, we consider the system  in the continuum limit (CL). For this reason, we  transform the main part of the sum in $\mathcal F(z)$ into an integral
 %%%%
 \begin{align}
   \sum_{k \neq 0    } \frac{ t_{k,\alpha}^2}{z + i \omega_{k,
 \alpha}}   &   \rightarrow    \frac 1{2\pi} \int_0^{\infty} d \omega' \frac{\Gamma_\alpha
 (\omega')}{z + i \omega'}  \equiv C_\alpha (z) ,
 \label{eq:continuum}
 \end{align}
  %%%%%
  with  the tunnel rate $\Gamma_\alpha (\omega)= 2\pi \sum_{k}  t_{k\alpha}^2 \delta(\omega -\omega_{k\alpha})$, which we assume to be analytic for $\omega>0$ in the CL.
   We emphasize that in order to  investigate the time evolution of the condensate, it is necessary to extract the ground-state energy modes $k=0$ from the integral.
  In the CL, we thus have
  \begin{equation}
 \mathcal  F(z) = z + i \epsilon +  \sum_{\alpha =L,R} \frac{ t_{0 ,\alpha}^2}{z + i \omega_{0 ,\alpha}}  +  C_\alpha (z) .
  \label{eq:characteristicFunctionGS}
  \end{equation}
  This approach is  a modification of former investigations of the Fano-Anderson model as in Ref.~\cite{Topp2015}. The extraction of the ground-state modes   allows for a detailed analysis of the condensate dynamics
 and  creates a three-mode system with modes $\myC_{0,L}$, $\myC_{0,R}$, and $\myD$ which is coupled to the reservoirs. The latter are described by the tunnel rates $\Gamma_\alpha(\omega)$. A sketch of the resulting setup is depicted in Fig.~\ref{fig:overview}(d).  
  In Sec.~\ref{sec:EffectiveHamiltonian}, we derive an effective non-hermitian Hamiltonian for the three-mode system which exactly resembles the dynamics in the long-time limit.
  The non-hermitian property  reflects the fact  that the coherent dynamics in this system is  subjected to loss and is thus not unitary. Our approach thus provides the possibility to analytically study these effects.
  
  We consider a   tunnel rate  with an exponential cut-off in its energy dependence, thus,
  \begin{equation}
  	\Gamma_\alpha (\omega) = \gamma_\alpha \left(\frac \omega{\omega_c}\right)^\eta  e^{- \frac \omega{\omega_c}} \Theta(\omega),
  	\label{eq:spectralDensity}
  \end{equation}
  where $\eta>-1$ is a scaling exponent describing the tunnel rate close to $\omega=0$. The parameter $\gamma_\alpha$ is the coupling constant and $\omega_c$ denotes the cut-off frequency.
  The function $\Theta(\omega)$ denotes the Heavyside function and guarantees that the reservoir spectrum is bounded at $\omega=0$.
   We choose this tunnel rate, as it allows for analytical calculations. However, many of our results as, e.g.,  the complex frequencies in Eq.~\eqref{eq:rootApprox} are expressed in terms of $\Gamma_\alpha (\omega)$ itself and thus hold for more general parametrizations than Eq.~\eqref{eq:spectralDensity}. In our  investigations, we find that the dynamics of the condensate is mainly determined by the tunnel rates near $\omega\gtrsim0$. So the exact details as the cut-off of the tunnel rate are not important for our qualitative results.
 For the tunnel rate~\eqref{eq:spectralDensity} the integration in \eqref{eq:continuum} can be performed exactly and we obtain 
  \begin{equation}
   C_\alpha (z) =-i \frac{ \gamma_\alpha}{2\pi} \left(- i\frac{z}{\omega_c}\right)^\eta e^{- i\frac{z}{\omega_c}} \tilde \Gamma(1+\eta)\tilde \Gamma\left(-\eta,- i\frac{z}{\omega_c}\right),
   \label{eq:genericContinuum}
  \end{equation}
  where  $\tilde \Gamma(x)$ and $\tilde \Gamma(x,z)$ denote the complete and incomplete Gamma functions, respectively. The incomplete Gamma function $\tilde \Gamma(-\eta,- i\frac{z}{\omega_c} )$  is characterized by a branch-cut discontinuity in the complex  plane running from $z=-i \infty$ to $z=0$. Also the prefactor $\left(-i z /\omega_c \right)^\eta$ contributes for non-integer $\eta$ to the branch cut.

The branch cut is not a specific property of the chosen parametrization in~\eqref{eq:spectralDensity}, but is a generic property of $C_\alpha(z)$ as the integration in \eqref{eq:continuum} runs over positive frequencies only. It occurs since the integrand in \eqref{eq:continuum} has a pole at $z=-i \omega$. In Sec.~\ref{sec:branchCutAppr}, we discuss how to handle these branch-cut discontinuities analytically.

 \subsection{Dynamical regimes}

\label{sec:DynamicalRegimes}

 We calculate the  time evolution of the system operators by performing an exact inverse Laplace transformation of \eqref{eq:SolutionLaplaceD} and \eqref{eq:SolutionLaplaceC}.
 The expectation values of the observables we are interested in, such as the current $I_{R,J}(t)=\text{tr} \left[\mathbf I_{R,J}(t) \rho \right]$, depend on correlation functions, e.g., $\left<\myC_{k,\alpha}^\dagger \myC_{k',\alpha'} \right>_0$ at $t=0$. The expectation value $\left< \mathbf O\right>_t$ is defined by
 \begin{equation}
	 \left< \mathbf O\right>_t\equiv \text{tr} \left[\mathbf O (t) \rho \right].
 \end{equation}
 For the initial condition~\eqref{eq:InitalCondition1} in the zero-temperature limit,  the only non-vanishing correlation functions are
  \begin{align} 
    \left< \myC_{k,\alpha}^\dagger \myC_{k,\alpha}\right>_0 &= \delta_{k,0} n_\alpha, \nonumber \\
    \left< \myC_{0,R}^\dagger \myC_{0,L}\right>_0 &= \sqrt{ n_L n_R} e^{i \Delta \phi},
    \label{eq:InitalCorrelation}
  \end{align} 
 where $\Delta\phi= \phi_R-\phi_L$ denotes the initial phase difference of the left and right condensate,
and $n_\alpha$ their initial occupation.
 
 The results are depicted in Fig.~\ref{fig:overview}(c) and   Fig.~\ref{fig:Current}. In the numerical calculations throughout the article  we  assume a symmetric system, meaning that $\omega_{0,L}=\omega_{0,R}=0$. In the time evolution we observe two dynamical regimes. They can be distinguished best by considering the dot occupation $N_d$, the Josephson current $I_{R,J}$, and the occupation of the right ground-state mode $N_{0,R}$.
 
  In regime I for $\epsilon> \epsilon_{crit} \approx 0.016 \omega_c$, we observe that $N_d(t)$ and $I_{R,J}(t)$ exhibit  initial oscillations that are quickly damped.
  The determination of $\epsilon_{crit}$ is discussed in Sec.~\ref{sec:RootsOfSymmeticSystem}. For longer times, we find that  $N_d(t)$ reaches a constant value while $I_{R,J}(t)$ performs  oscillations  with a very long period, which are undamped. The oscillation frequency increases when approaching $\epsilon=\epsilon_{crit}$. During these oscillations the main fraction of the particles oscillates between the two reservoir ground states. A rather small amount of the initially condensed particles are subjected to depletion. They are scattered to the excited modes during the dynamics (difference between the curves in Fig.~\ref{fig:Current}(b)). The depletion is stronger for $\epsilon$ close to $\epsilon_{crit}$.

 In regime II for $\epsilon< \epsilon_{crit}$, the dot occupation $N_d(t)$   exhibits   fast  oscillations  which are only weakly damped. The damping is stronger close to the  transition. The time evolution of  $I_{R,J}(t)$ displays a superposition of two oscillations with long and short period, respectively. However, only the slow oscillations significantly change the occupation of the right reservoir in panel (b). The depletion is only noticeable close to $\epsilon_{crit}$.
 
 In   Fig.~\ref{fig:Current}(c), we depict a finite-size simulation of the Josephson current as a benchmark for our approach. In appendix~\ref{sec:AppFiniteSizeSimulation}, we provide  information about its calculation. Thereby, each reservoir consists of a rather small number of modes, namely $k_{max}=100$. 
 
  As we observe in  Fig.~\ref{fig:Current}(c), the numerical finite-size simulation agrees well with the CL calculations for short times $t \omega_c \cdot 10^3\lesssim 1 $ after which the finite-size simulation starts to exhibit deviations. These appear due to the finite energy spacing $\delta \omega \approx  \omega_c/k_{max}$ between the levels close to the ground state. In  numerical studies we find that the starting time of these deviations $T_{dev}$ grows  for increasing $k_{max}$.

  However, we assume that these fluctuations are not particularly relevant in  experiments. The actual observable quantity is the particle number in the right reservoir which is mainly given by the condensate particles $N_{0,R}$. As we see in Fig.~\ref{fig:Current}(d), the fluctuations in the current are averaged so that they are hardly visible in $N_{0,R}$. Additionally, weak interactions which are always present in experiments could induce a damping of these finite-size fluctuations.
  
 Finally, we emphasize that the two regimes I and II are connected by a smooth crossover as a function of $\epsilon$ for finite $t_{0,\alpha}$. This becomes clear when considering the relation of observables and roots of the system later in this article in Sec.~\ref{sec:RelationObsRoots}. The transition is only non-analytic in the limit $t_{0,\alpha}\rightarrow0$.

 \section{Long-time limit}
 
  \label{sec:LongTimeLimit}
  
    %%%
    %%%
    \begin{figure*}[t]
      \centering
      \includegraphics[width=1\linewidth]{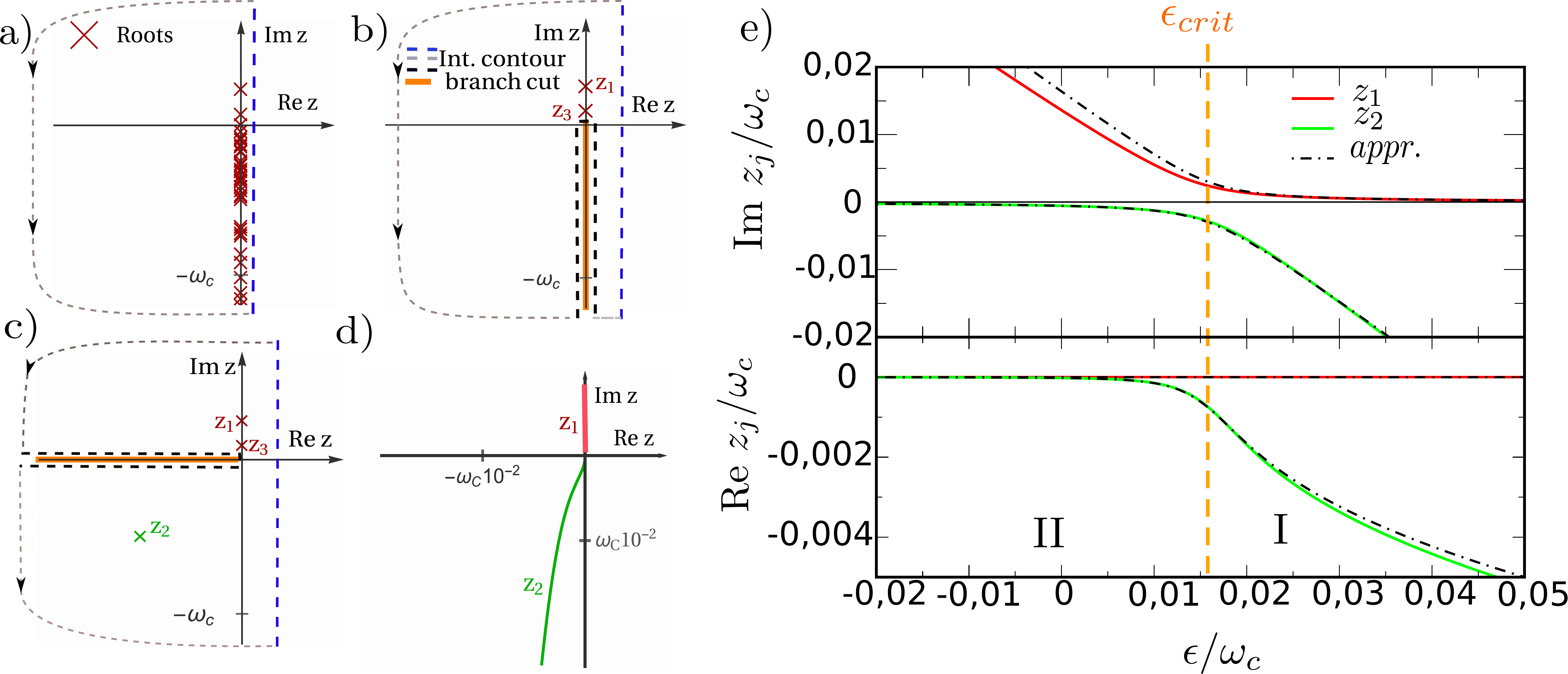}
      \caption{(Color online) \textbf{(a)} Sketch of the roots of $\mathcal F(z)$ in Eq.~\eqref{eq:characteristicFunction} for a finite-system with $k_{Max}=100$. The dashed curve depicts the contour used to apply the residue theorem. See main text for more details. \textbf{(b)} In the CL, the roots get dense and finally form a branch cut (orange) located at the negative imaginary axes. In addition, there are two roots $z_1$ and $z_3$ generated by the extraction of the ground-state modes in Eq.~\eqref{eq:characteristicFunctionGS}. The black dashed curve depicts the integration contour $\mathcal C$ needed to evaluate Eq.~\eqref{eq:InverseLaplace}. \textbf{(c)} Rotation of the branch cut so that its contribution in the inverse Laplace transformation~\eqref{eq:InverseLaplace} is negligible in the long-time limit. Due to the rotation, the modified $ \mathcal {\tilde F}(z)$ given in~\eqref{eq:modifiedF} exhibits an additional root with finite real part $z_2$.  
      \textbf{(d)} Trajectories of $z_1$ and $z_2$ in the complex plain as function of $\epsilon$. The parameters are identical to Fig.~\ref{fig:overview}(c) so that  $z_3\rightarrow 0$.
      \textbf{(e)} Real and imaginary part of the roots $z_1$ and $z_2$ as a function of $\epsilon$. The dashed line depicts the approximation \eqref{eq:rootApprox}. 
      }
      \label{fig:roots}
      \end{figure*}
    %%%
    %%%
 
 \subsection{Analysis in the complex plain}
 
    \label{sec:branchCutAppr}
 
In order to better understand the time evolution in Fig.~\ref{fig:Current}, we investigate the time evolution for intermediate and long times in more detail. 
To this end, we  have to identify the main contributions in the inverse Laplace transformation of~\eqref{eq:SolutionLaplaceD} and~\eqref{eq:SolutionLaplaceC}. In particular, the analytic properties of $\mathcal F(z)$ and its roots are important for the inverse Laplace transform so that we analyze it in the following. 

 First, we consider the roots of $\mathcal F(z)$ in~\eqref{eq:characteristicFunction}   for a finite-sized system with $k_{max}$
states in the reservoirs. The roots of $\mathcal F(z)$ are located on the imaginary axis as sketched in Fig.~\ref{fig:roots}(a).
Using the residue theorem with the dashed contour shown in Fig.~\ref{fig:roots}(a), we find that one can perform the inverse Laplace transformation   
\eqref{eq:InverseLaplaceTrafo} for $ \mathbf o =   \myD,\myC_{k\alpha} $ corresponding to the blue integration contour, by evaluating the residua 
  \begin{equation}
   \mathbf o(t) 	= \sum_{a\in D_{fs} }   \text{Res}_{z=a} \quad  e^{z t} \hat {\mathbf{o}}(z),
   \label{eq:InverseLaplaceFS}
  \end{equation}
where $D_{fs}$ is the set of all poles in~\eqref{eq:SolutionLaplaceD} and~\eqref{eq:SolutionLaplaceC}
which includes  the set of all roots of $\mathcal F(z)$. 
  
 For $k_{max}\rightarrow \infty$, the roots move closer
to each other and finally form the branch cut of $\mathcal F(z)$ in~\eqref{eq:characteristicFunctionGS}. The branch cut is depicted in Fig.~\ref{fig:roots}(b). Everywhere else in the complex plane, $\mathcal F(z)$ is analytic. The branch cut of $\mathcal F(z)$  is due to the branch cut of $C_\alpha(z)$, c.f. Eq.~\eqref{eq:genericContinuum}.
At the branch cut,  the function $C_\alpha(z)$ has a jump discontinuity of
\begin{equation}
  \lim_{\delta\downarrow 0 } \; C_\alpha(-i \omega+\delta)- C_\alpha(-i \omega-\delta) = \Gamma_\alpha(\omega) ,
  \label{eq:JumpOfC}
\end{equation}
where $\Gamma_\alpha(\omega)$ is the tunnel rate and $\omega>0,\delta \in \mathbb R$. This relation can be proven using Eq.~\eqref{eq:continuum} and the Dirac identity
\begin{equation} 
	\lim_{\delta\downarrow 0}\frac{1}{(\omega-\omega')\pm i \delta} = P \frac{1}{\omega-\omega'} \mp i \pi \delta(\omega-\omega'),
	\label{eq:DiracIdentity}
\end{equation}
where $P$ denotes the principal value.

 Besides, for the chosen parameters in Fig.~\ref{fig:roots}(b) there are two roots $z_1$ and $z_3$ which are not merged with the branch cut. They appear due to the extraction of the ground states as explained in Eq.~\eqref{eq:characteristicFunctionGS}.

As for the finite-size system, we  apply the residue theorem with the dashed contour shown in Fig.~\ref{fig:roots}(b) in order to evaluate the inverse Laplace transformation~\eqref{eq:InverseLaplaceTrafo}. The integration contour is chosen  so that the surrounded area is analytic except of isolated poles.  In doing so, we find
  \begin{equation}
   \mathbf o(t) 	= \sum_{a\in D }   \text{Res}_{z=a} \quad  e^{z t} \hat {\mathbf{o}}(z)  -  \frac1{2\pi i}  \int_{ \mathcal C} e^{z t}  \mathbf{o}(z) dz ,
   \label{eq:InverseLaplace}
  \end{equation}
where  $\mathcal D $ is the set of all isolated poles appearing in~\eqref{eq:SolutionLaplaceD} and~\eqref{eq:SolutionLaplaceC} which includes the roots of $\mathcal F(z)$ in~\eqref{eq:characteristicFunctionGS}. Formula \eqref{eq:InverseLaplace} is valid under the assumption of a vanishing integrand for $\text{Re}\;z \rightarrow - \infty$. For this reason, we can omit the gray integration contour in Fig.~\ref{fig:roots}(b). The only remaining integration contour is $\mathcal C$   depicted in black in Fig.~\ref{fig:roots}(b) and encircles the branch cut of $\mathcal F(z)$.
Equation~\eqref{eq:InverseLaplace}  constitutes an exact expression in the CL. However, the evaluation of the branch-cut integral in Eq.~\eqref{eq:InverseLaplace} is numerically expensive and analytically unfavorable. For this reason, we explain in the following how to circumvent this problem in the long-time limit.

The branch cut of $C_\alpha(z)$ in \eqref{eq:continuum} which generates the branch cut of $\mathcal F(z)$ is not uniquely defined. There is the possibility to modify $C_\alpha(z)$  so that its branch cut is located elsewhere. As the branch cut separates the bottom-left and the bottom-right sector of the complex plane, we modify $C_\alpha(z)$ in the bottom-left sector, so that the current branch cut is displaced and the modified function $\tilde C_\alpha(z)$ is analytic on the negative imaginary axis. More, precisely, that modification reads
 \begin{equation}
 \tilde C_\alpha(z)  =  C_\alpha(z) +
	\begin{cases}
		 \Gamma_\alpha(i z) & \text{Re}\;z <0 \land \text{Im}\; z <0 \\
		0 & \text{else}
	\end{cases},
	\label{eq:branchCutRotated}
 \end{equation} 
 where $ \Gamma_\alpha(i z) $ is the analytic continuation of the tunnel rate as defined for $\omega=iz>0$  onto the complex plane. One can find from Eq.~\eqref{eq:JumpOfC} that $\tilde C_\alpha(z)$ is continuous along the previous branch cut position.
In appendix \ref{sec:AppendixBranchCutRotation} we prove, that $\tilde C_\alpha(z)$ is indeed analytic on the negative imaginary axis. In addition, as $\tilde C_\alpha(z)$ is a sum of analytic functions in the bottom-left sector, it is  analytic there. Yet, due to the modification, $\tilde C_\alpha(z)$ is not continuous on the negative real axis separating the top-left and bottom-left sector as depicted in Fig.~\ref{fig:roots}(c). Consequently,  $\tilde C_\alpha(z)$ has now a branch cut there. Thus, $ C_\alpha(z)$ and $\tilde C_\alpha(z)$ are related by a branch cut rotation.
 
   Consequently, we also  modify
   \begin{equation}
   \mathcal{\tilde F}(z)\equiv \mathcal{ F}(z) - \sum_{\alpha=L,R}   C_\alpha(z) + \sum_{\alpha=L,R}  \tilde  C_\alpha(z),
   \label{eq:modifiedF}
   \end{equation}
   which is therefore also  analytic everywhere except on the negative real axis.
As $\mathcal{\tilde F}(z)= \mathcal{ F}(z)$ for $\text{Re}\; z >0$, the inverse Laplace transformation in Eq.~\eqref{eq:InverseLaplaceTrafo} is not affected so that the operators as a function of time remain invariant under the branch-cut rotation. As before, we employ the residue theorem to simplify the evaluation of the inverse Laplace transformation in Eq.~\eqref{eq:InverseLaplaceTrafo} with the blue integration contour in Fig.~\ref{fig:roots}(c).  The corresponding contour is depicted by the dashed lines in Fig.~\ref{fig:roots}(c). The result is formally equivalent to Eq.~\eqref{eq:InverseLaplace}, but with the integration contour $\mathcal C$ depicted in black in Fig.~\ref{fig:roots}(c).

    $\mathcal{\tilde F}(z)$ is different from $ \mathcal{ F}(z)$ in the third sector of the complex plain. This gives rise  to
    an additional root $z_2$ with negative real part.   We depict it also in Fig.~\ref{fig:roots}(c). In the limit of $t_{k,\alpha}\rightarrow 0$, it corresponds to the
    gate potential $\epsilon= i z_2$ in regime I. 
    For the symmetric system, we calculate the leading orders of the position of the root for small $t_{\alpha,0}$ in appendix~\ref{sec:AppendixRootsDerivation}.
    Altogether, $ \mathcal{\tilde F}(z)$ possesses three roots. We found  that the number of roots of $ \mathcal{\tilde F}(z)$  is independent of the system parameters.
    
    We approximate now the inverse Laplace transformation by neglecting the branch-cut integral in Eq.~\eqref{eq:InverseLaplace}. This is justified as the  integrand in \eqref{eq:InverseLaplace} contains the factor $ e^{z t}$, which vanishes in the long-time limit as the integration contour surrounds the negative real axis.
    For example,  the branch-cut integral $\mathcal I_{bc}$ belonging to the  second line in Eq.~\eqref{eq:SolutionLaplaceC} for $k,k'=0$  is for long times approximately given by 
      \begin{equation}
   	   \mathcal I_{bc} \approx \frac1{2\pi i}   \frac {  \sum_{\alpha} \gamma_\alpha    }{4 t_0^2  } \left(\frac{-i}{\omega_c}\right)^\eta\tilde \Gamma(\eta+1)  \frac 1{ t^{\eta+1}  } c_{0,\alpha'},
   	   \label{eq:branchCutAppr0}
      \end{equation}
      if $\omega_{0,\alpha}= \omega_{0,\alpha'}=0$ and $t_{0,L}=t_{0,R}=t_{0}$ which is the most important case in our article. In  appendix~\ref{sec:AppBranchCutApproximation}, we provide more information about the calculation. There, we also consider the cases  $\omega_{0,\alpha}\neq \omega_{0,\alpha'}=0$ and $\omega_{0,\alpha}\neq 0\neq \omega_{0,\alpha'}$ which yield similar results. 
     From \eqref{eq:branchCutAppr0} we see, that the branch cut integral vanishes algebraically in time for long times if $\eta>-1$. Branch-cut integrals corresponding to the other terms in~\eqref{eq:SolutionLaplaceD} and ~\eqref{eq:SolutionLaplaceC} vanish even faster.
     
  For intermediate times we found that  the branch-cut integral contributes only insignificantly for $\epsilon$ away from the transition point  $\epsilon_{crit}$. As a result, the dynamics of the system for intermediate and long times  is  determined by the poles of \eqref{eq:SolutionLaplaceD} and \eqref{eq:SolutionLaplaceC}, which we can efficiently calculate numerically.
  
  Finally, we have to point out a subtlety. The function $\mathcal{\tilde F}(z)$ exhibits only three roots for  $\omega_{0,L}\neq \omega_{0,R}$. For the special case  $\omega_{0,L}=\omega_{0,R}$ which we mainly focus on in this article, $\mathcal{\tilde F}(z)$ has indeed only two roots. The missing root corresponds to a dark state with  energy $\omega_{0,L}$.
  The corresponding mode reads
  \begin{equation}
  \myC_{dark} = \frac 1{\sqrt {t_{0,L}^2+t_{0,R}^2} } \left(t_{0,L}\myC_{0,R} - t_{0,R}\myC_{0,L}\right).
  \label{eq:DarkMode}
  \end{equation}
  This mode obviously does not couple to the dot or the excited reservoir modes.  For this reason, it does not appear in $\mathcal{\tilde F}(z)$.
  It is not hard to show that $\left[\mathbf H ,\myC_{dark}^\dagger\myC_{dark} \right]=0$.
  Therefore, the particle number in the dark mode remains constant and it is not subjected to particle loss. Consequently, there is no complete depletion of the ground-state modes  if the dark state is initially occupied.
   In our generic investigation in Sec.~\ref{sec:EffectiveHamiltonian}   we consider the more general case of $\omega_{0,L}\neq \omega_{0,R}$ and regard the equality as the  limit $\omega_{0,L}\rightarrow \omega_{0,R}$. 
   
   Furthermore, if additionally $\omega_{0,L}\rightarrow \omega_{0,R}=0$, than the dark-state root $z_3\rightarrow0$, which lies within the branch-cut contour $\mathcal C$. For this reason, we  treat this special case formally with a limit procedure: first we assume  $\omega_{0,L}\rightarrow\omega_{0,R}\neq0$ and than take the limit $\omega_{0,R}\rightarrow0$ after performing the inverse Laplace transformation.
  
  \subsection{Roots of the symmetric system}
  
  \label{sec:RootsOfSymmeticSystem}
  
  In the following, we analyze the roots of $\mathcal{\tilde F}(z)$ for the symmetric system with $\omega_{0,L}=\omega_{0,R}=0$. For simplicity we also assume symmetric tunneling rates $t_{0,L}= t_{0,R}\equiv t_0$ . Here, we give an analytical expression for the leading contributions of the real and imaginary part of the roots $z_1$ and $z_2$.

In order to express the location of the roots, we define the real and the imaginary part of $\sum_{\alpha}\tilde C_\alpha(z)$ by
\begin{equation}
	\lim_{\delta\downarrow 0}\sum_{\alpha =L,R} \tilde C_\alpha(-i \omega + \delta) \equiv \Gamma(\omega) + i \Sigma(\omega),
	\label{eq:DampingAndLampshift}
\end{equation}
with $\omega,\delta\in \mathbb R$.
The real part can be expressed with the tunnel rates
\begin{equation}
	 \Gamma(\omega) = \frac 12 \left(\Gamma_L(\omega)+\Gamma_R(\omega) \right).
\end{equation}
which can be proven using the Dirac identity~\eqref{eq:DiracIdentity}.

In  appendix~\ref{sec:AppendixRootsDerivation} we show that for small $t_0$ and $\eta>0$ the roots $z_1$ and $z_2$ of $\mathcal{\tilde F}(z)$ are approximately located at
\begin{align}
	z_j \approx z_j^0 -  \frac {z_j^0  \Gamma( i z_j^0) }{  2 z_j^0 + i\left[ \epsilon + \Sigma(0)\right] },
	\label{eq:rootApprox}
\end{align}
where  the imaginary part  $z_j^0$ reads
\begin{align}
	 z_{1,2}^{0} &= - i\frac 1 {2 } \left(\epsilon + \Sigma(0) \pm \sqrt{\left( \epsilon + \Sigma(0)\right)^2 + 8 t_0^2 }\right) .
	 \label{eq:ImRootApprox}
\end{align}
 Using Eq.~\eqref{eq:genericContinuum} we  find for   the Lamb shift
\begin{equation}
	 \Sigma(0) = - \sum_{\alpha=L,R}\frac{\gamma_\alpha}{2\pi}  \tilde \Gamma(\eta) ,
	 \label{eq:LampShift}
\end{equation}
 which renormalizes the gate potential $\epsilon$.
The second term in Eq.~\eqref{eq:rootApprox} is the leading order of the real part. Interestingly, it is proportional to $\Gamma( i z_j^0)$. Consequently, if $i z_j^0<0 $,  the real part vanishes due to Eq.~\eqref{eq:spectralDensity}.

The  analytical expressions for $z_1$ and $z_2$ are depicted in Fig.~\ref{fig:roots}(e) and agree well with the numerical calculation. The imaginary parts of the  roots  $z_j$ can be used to define 
the transition between the two dynamical regimes I and II. 
For $t_0\rightarrow 0$, the two roots get degenerate for 
\begin{equation}
	\epsilon=\epsilon_{crit}\equiv - \Sigma(0).
\end{equation}
This relation defines the regime I for $\epsilon>\epsilon_{crit}$ and regime II for $\epsilon<\epsilon_{crit}$.
As we see in  Fig.~\ref{fig:roots}(e), the root $z_2$  has the property $\text{Re}\;z_2<0$. In the  regime II we have $\text{Re}\;z_2\approx0$ which vanishes exactly for $t_0=0$ as we can see from~\eqref{eq:rootApprox}. Yet, in regime I it is always finite. By contrast, the real part of $z_1$ is always $\text{Re}\; z_1=0$. To clarify the dependence of $z_1$ and $z_2$ on $\epsilon$, we also depict the trajectory of these roots in the complex plain as a function of $\epsilon$ in Fig.~\ref{fig:roots}(d).

\subsection{Effective Hamiltonian}
\label{sec:EffectiveHamiltonian}

We are  interested in the  dynamics of observables which can be expressed by  the operators $\myD$, $\myC_{0,L}$, and $\myC_{0,R}$, such as the dot occupation or  the Josephson current.  As we have assumed a zero-temperature limit at time $t=0$, the only relevant operators in Eqs.~\eqref{eq:SolutionLaplaceD} and \eqref{eq:SolutionLaplaceC} at $t=0$ are even these operators. For this reason, we can effectively restrict the BEC dynamics to a three-mode system.

 For a notational reason  we define
\begin{equation}
	\mathbf v_t\equiv  
	\left(\begin{array}{c}
	\myD(t)\\
	\myC_{0,L}(t)\\
	\myC_{0,R}(t)
		\end{array}\right).
		\label{eq:DefV}
\end{equation}
The dynamics of $ \mathbf v_t$ in the long-time limit is determined by an effective non-hermitian Hamiltonian, which fulfills
\begin{equation}
	i\frac{d}{dt}\mathbf v_t = \mathcal H_{eff}  \mathbf v_t,
	\label{eq:effSchroedingerEq}
\end{equation}
which is formally equivalent to a single-particle Schr\"odinger equation.
Thereby, the effective Hamiltonian $\mathcal H_{eff} $ reads
\begin{equation}
\mathcal H_{eff} = S D S^{-1} ,
\end{equation}
where $D= \text{diag}\left[i z_1,i z_2, i z_3 \right]$ is a diagonal matrix containing the roots of $\mathcal {\tilde F}(z)$. The columns of the matrix $S$ are  given by $S_j =w_j $ with
\begin{equation}
	w_j= \frac1{\sqrt{\zeta_j}}\left(-1,  \frac{t_{0,L}}{( \omega_{0,L}-i z_j)} ,   \frac{t_{0,R}}{( \omega_{0,R}-i z_j) } \right)^T,
	\label{eq:EigenstateQ}
\end{equation}
where  $\zeta_j $  accounts for the normalization. We note that the $w_j$ are in general not orthogonal to each other. Additionally, the roots $z_j$ can be complex-valued so that  the effective Hamiltonian is  non-hermitian. 

We remark that the only important poles $z_i$ for the effective Hamiltonian are the ones given by $\mathcal {\tilde F}(z)=0$. The other poles appearing in \eqref{eq:SolutionLaplaceD} and \eqref{eq:SolutionLaplaceC} such as $z=-i \omega_{0,\alpha} $ are not relevant. More precisely, the factors $ (z+ i\omega_{0,\alpha} )$ appearing in the nominators can be combined with
 $\mathcal {\tilde F}(z)$, which cancels  the nominators in the third term of \eqref{eq:characteristicFunctionGS}. This combination therefore reveals that the first order pole at $z=- i\omega_{0,\alpha}$ is not an actual pole in second term of \eqref{eq:SolutionLaplaceD} and in the second and third term of \eqref{eq:SolutionLaplaceC}. Finally, the pole of the first term in \eqref{eq:SolutionLaplaceC} at $z=-i \omega_{0,\alpha}$ is annihilated by the term with $\left(z+i \omega_{0,\alpha}\right)^2 $ in the nominator of the third term in \eqref{eq:SolutionLaplaceC} during the inverse Laplace transformation.

We note that  effective non-hermitian  Hamiltonians can give rise to interesting effects not present in hermitian systems. A particular appealing effect is a non-hermitian degeneracy, at which two eigenvalues and their corresponding eigenstates merge which is denoted as an exceptional point~\cite{Berry2004,Heiss2012,Gao2015,Doppler2016}. Usually, the construction of  non-hermitian Hamiltonians includes phenomenological assumptions. Here, we presented a completely microscopic derivation which can be used to study the fate of exceptional points under more realistic conditions. In particular, here the eigenvalues are not determined by the roots of the characteristic polynomial of the Hamiltonian, but by the roots of  $\mathcal{\tilde F}(z)=0 $, which exhibits a non-linearity due to $\tilde C_\alpha (z)$. This might gives rise to a qualitatively different behavior of the system at or close to the exceptional points.

In the remainder of this section we prove that $\mathcal H_{eff} $ generates indeed the correct dynamics in the long-time limit.
After neglecting the branch cut in Eq.~\eqref{eq:InverseLaplace}, the time evolution can be evaluated by calculating the corresponding residua at the roots of $\mathcal{\tilde F}(z)$.
In doing so, the  time evolution   of the operators reads
\begin{equation}
	\mathbf v_t = \sum_{z_j \in \mathcal D_{\mathcal F}}  e^{z_j t}    Q(z_j) \mathbf v_0,
	\label{eq:TimeEvolution operator}
\end{equation}
where we define $\mathcal D_{\mathcal F}= \left\lbrace z \mid \mathcal{\tilde F}(z)=0\right\rbrace $  which is the set of all three roots of $\mathcal{\tilde F}(z)=0 $. Here $Q(z_j)$ denotes a $3\times3$ matrix
and reads
\begin{align}
	&Q( z_j) = R_{z_j} \times  \label{eq:QMatrix}\\
	&\left(
		\begin{array}{ccc}
				1                                   & \frac{-t_{0,L}}{( \omega_{0,L}-i z_j)} &  \frac{-t_{0,R}}{( \omega_{0,R}-i z_j)} \\
		 \frac{-t_{0,L}}{( \omega_{0,L}-i z_j)}& \frac{t_{0,L}^2}{( \omega_{0,L}-i z_j)^2} & \frac{t_{0,R}t_{0,L}}{( \omega_{0,R}-i z_j)( \omega_{0,L}-i z_j)} \\
	  \frac{-t_{0,R}}{( \omega_{0,R}-i z_j)}&\frac{t_{0,R}t_{0,L}}{( \omega_{0,R}-i z_j)( \omega_{0,L}-i z_j)} & \frac{t_{0,R}^2}{( \omega_{0,R}-i z_j)^2}  \nonumber
		\end{array}
	\right)
\end{align}
with
\begin{equation}
 R_{z_j} = \text{Res}_{z= z_j} \frac 1{\mathcal {\tilde F}(z)}.
\end{equation}

The matrix $ Q(z_j)$ is hermitian  for imaginary $z_j$. Interestingly, it fulfills a projector-like relation
\begin{align}
	 Q(z_j) Q(z_j)&= Q(z_j) \;  \zeta_{z_j} R_{z_j}  .
\end{align}
This relation even holds for complex-valued $z_j$.
 Consequently, two eigenvalues of $ Q(z_j)$ are zero. The non-vanishing eigenvalue is $\zeta_{z_j} R_{z_j} $. For $\gamma_\alpha=0$ we have a bare three-mode system without coupling to the excited modes. For this reason, the eigenvalue  is necessarily $\zeta_{z_j} R_{z_j}=1$, so that the time evolution is unitary. Due to the coupling  to the excited reservoir modes for $ \gamma_\alpha>0 $, it is possible that $\zeta_{z_j} R_{z_j} \neq 1$. This eigenvalue thus includes information about the transient dynamics. 

The normalized eigenstates corresponding to $ R_{z_j}\zeta_{z_j}$ are the  $w_j$ given in Eq.~\eqref{eq:EigenstateQ}.
Consequently, the matrix  $ Q(z_j)$ can be written as
\begin{equation}
	Q(z_j)= \zeta_{z_j} R_{z_j} \; w_j    w_j^T  .
\end{equation}
Inserting this into Eq.~\eqref{eq:TimeEvolution operator} and using that $ w_j $ is an eigenstate of $\mathcal H_{eff}$, we can verify that Eq.~\eqref{eq:TimeEvolution operator} fulfills the Schr\"odinger equation \eqref{eq:effSchroedingerEq}.

\subsection{Relation of observables and roots}

\label{sec:RelationObsRoots}

The properties of the roots are reflected in the oscillations of the observables.
  For  example, the Josephson current for long times reads
    \begin{equation}
    	 I_{R,J} (t)    =  2 \text{Re} \sum_{j,j'}e^{(z_j^*+z_{j'})t}  \tilde  I_{R,J}^C  (z_j,z_{j'}) ,
    	\label{eq:Current}
    \end{equation}
where the constants $ I_{R,J}^C  (z_j,z_{j'})$ are a function of the roots and depend on the initial condition. Their explicit expressions can be found in Eq.~\eqref{eq:CurrentFourier2}. The time evolution of other observables such as the ground-state occupation of the right reservoir $N_{0,R}(t)$ read similarly. We 
see that the oscillations are determined by the exponential factor $e^{(z_j^*+z_{j'})t}  $. We find  that  $ I_{R,J}^C  (z_j,z_{j})=0$. For the symmetric system $\omega_{0,\alpha}=0$ and $t_{0,\alpha}=t_0$,   $ I_{R,J}^C  (z_j,z_{j'})$ is rather small if both $z_j,z_{j'}\neq z_3 =0$. So the most important terms are the ones where one root is $z_1$ or $z_2$ and the other is $z_3=0$. Thus, the dynamics is mainly determined by the roots $z_1$ and $z_2$.
  Thereby, the imaginary parts are responsible for the oscillation frequencies while the real parts determine the damping.
  The oscillatory behavior of other observables  such as the ground-state occupation of the right reservoir $N_{0,R}(t)$ or the dot occupation $N_d(t)$ is similarly determined by the roots $z_1$ and $z_2$.
  
    As a consequence of the finite real part of $z_2$ in regime I, the fast oscillations  in the observables caused by the imaginary part of $z_2$ are strongly damped as can be seen in Fig.~\ref{fig:Current}(a)  for $\epsilon = 0.04 \omega_c$. The oscillations of $I_{R,J}$ with a long period are caused by $z_1$, as it has a small imaginary part. They are undamped as  $\text{Re}\;z_1=0$.
    
   In regime II, the imaginary part of $z_1$ is large. For this reason we observe fast oscillations which are undamped. 
   The imaginary part of $z_2$ is small so that it causes oscillations with a long period.
 However, there is a very small damping due to the very small real part of $z_2$. This can be seen best in  $N_{0,R}(t)$ in Fig.~\ref{fig:Current}(b) for $\epsilon=0$.
 
 Due to the  damping describing the particle loss in the condensate, the excited reservoir modes get occupied. 
As  $\text{Re}\;z_2$ is quite large in regime I, this condensate depletion is completed after a rather short time as can bee seen in Fig.~\ref{fig:Current}(b) for $\epsilon=0.02\omega_c$. By contrast, due to a small $\text{Re}\;z_2$ in regime II, the damping continues even for long times, so the fraction of the particles in the excited modes keeps growing as can be observed in  Fig.~\ref{fig:Current}(b) for $\epsilon=0$ and    $\epsilon=-0.02\omega_c$. We recall that  the particles which have been initially in the dark mode~Eq.~\eqref{eq:DarkMode} are not subjected to depletion. 

As the roots are a smooth function of the gate potential $\epsilon$ for finite $t_0$, c.f.Eq.~\eqref{eq:ImRootApprox} and Fig~\ref{fig:roots}(e), and as the time dependence of the observables is closely related to the roots, the crossover in the time evolution of the observables from regime I to regime II is also smooth. Only in the limit $t_0\rightarrow 0$ the crossover in the dynamics gets non-analytic which we used to define $\epsilon_{crit}$ in Sec.~\ref{sec:RootsOfSymmeticSystem}.

\subsection{Low-frequency current}

\label{sec:LowFrequencyCurrent}

    %%%
    %%%
    \begin{figure}[t]
      \centering
      \includegraphics[width=1\linewidth]{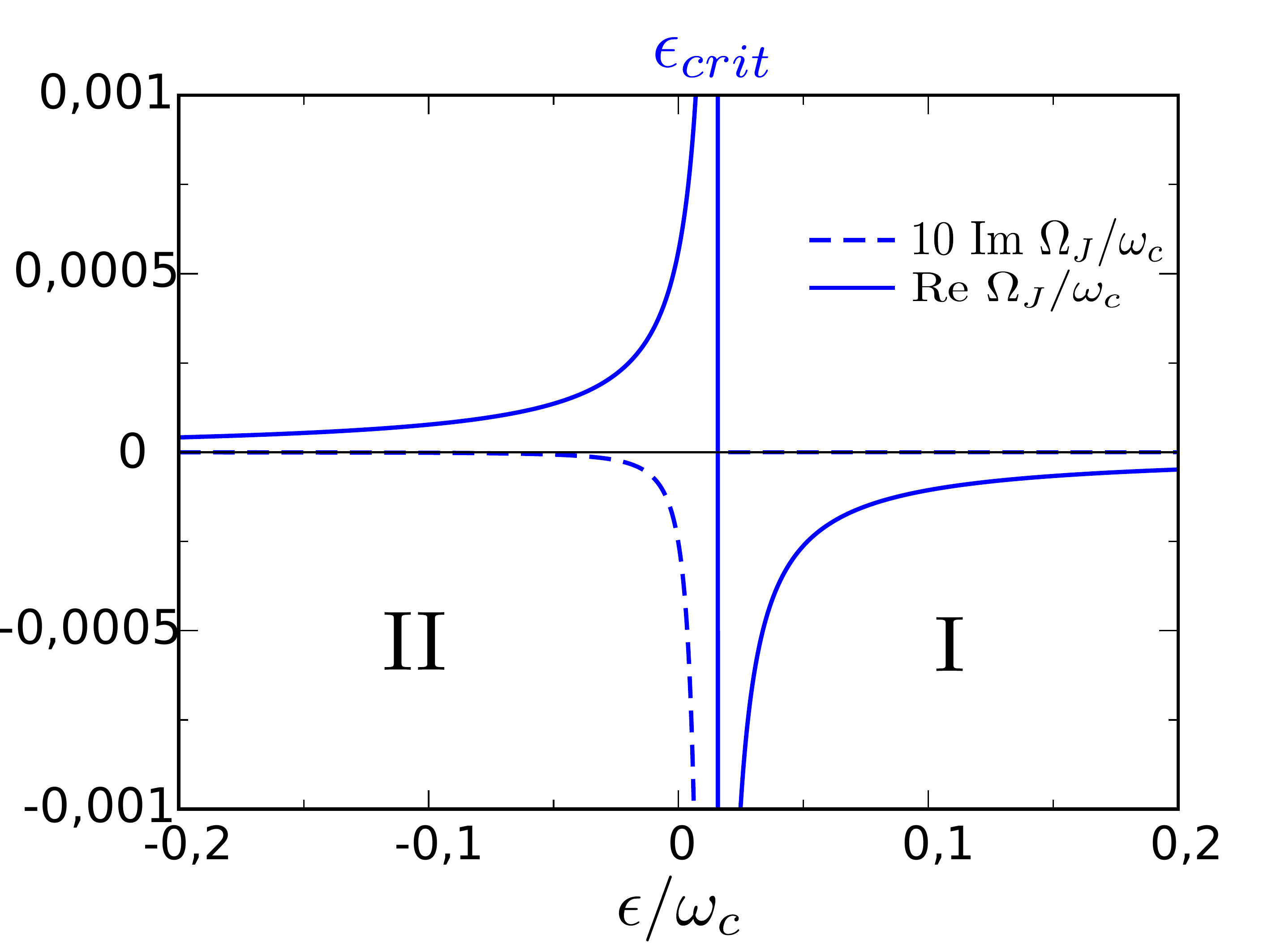}
      \caption{(Color online) Characteristic frequency $\Omega_J$ appearing in the low-frequency Josephson current Eq.~\eqref{eq:CurrentApprox}. The parameters are the same as in Fig.~\ref{fig:Current}.  In the immediate vicinity of $\epsilon_{crit}$, the expression \eqref{eq:CurrentApprox} is not valid as the oscillation frequencies of the Josephson current are of the same order.
      }
      \label{fig:characteristicFrequency}
      \end{figure}
    %%%
    %%%

As we can observe in Fig.~\ref{fig:Current}(b), the fast oscillations in the current  in panel (a) in regime II are averaged in time and induce only small variations in the particle number $N_{0,R}(t)$. For this reason, we  investigate the time-averaged dynamics in the following. 
Here we return to  the symmetric system, meaning that $\omega_{k,L}\rightarrow \omega_{k,R}=0$ and $t_{0,L}= t_{0,R}\equiv t_0$.  

The Josephson current for long times can be expanded in terms of its frequency contributions, c.f. Eq.~\eqref{eq:Current}. We define the low-frequency current by keeping only the contribution with the smallest frequency.
Formally, this corresponds to a moving time averaged with a time window $\tau$ around a time $t$. The duration $\tau$ has to be chosen so that $2\pi/\tau$ is smaller than all frequencies except one.
As we see in Fig.~\ref{fig:Current}(a) for $\epsilon=0$ and $\epsilon=-0.02\omega_c$, this is possible as the oscillation frequencies differ considerably away from the transition at $\epsilon=\epsilon_{crit}$.

In the  limit of small $t_0$, the low-frequency Josephson current  approximately reads 
\begin{align}
	\overline I_{R,J}(t)  &\approx  \frac 12 \text{Im}\; \left[  \Omega_J e^{-i \Omega_J t} \right]\left( n_R - n_L\right) \nonumber \\
		&+ \text{Re}\;\left[ \Omega_J e^{-i \Omega_J t} \right] \sin \Delta\phi \sqrt{n_L n_R},
		\label{eq:CurrentApprox}
\end{align}
where
\begin{equation}
	\Omega_J = \begin{cases}
						i z_1 & \epsilon>\epsilon_{crit}  \\
						i z_2 & \epsilon<\epsilon_{crit} 
				\end{cases}
				\label{eq:CharacteristicFrequecy}
\end{equation}
denotes the characteristic frequency. Thus, the low-frequency current is determined by the root $z_j$ that has the smaller imaginary part, c.f. Fig.~\ref{fig:roots}(e). 
Details of the derivation can be found in appendix~\ref{sec:AppendixTimeAveragedCurrent}. For $\epsilon\approx \epsilon_{crit}$, Eq.~\eqref{eq:CurrentApprox} fails as there is no clear separation of oscillation frequencies because $\text{Im}\; z_1$ and $\text{Im}\; z_2$ are in the same order of magnitude, c.f. Fig.~\ref{fig:roots}(e). For comparison, we have included the analytic expression~\eqref{eq:CurrentApprox} in Fig.~\ref{fig:Current}(a). We observe that it indeed resembles the time-averaged current away from $\epsilon_{crit}$.

Relation~\eqref{eq:CurrentApprox} is strongly reminiscent  of Eq.~\eqref{eq:Current2mode}. 
In contrast, here the characteristic frequency $\Omega_J$ is complex valued in regime II due to the depletion.
In regime I, $\Omega_J$ is purely real-valued so that Eq.~\eqref{eq:CurrentApprox} exactly resembles Eq.~\eqref{eq:Current2mode}.

As we can see in Eq.~\eqref{eq:CurrentApprox}, the characteristic frequency $\Omega_j$ determines the dynamics of the current.  We observe in Fig.~\ref{fig:characteristicFrequency} that the imaginary part of $\Omega_J$, is considerably smaller than the real part. Thus, the Josephson current is essentially proportional to the latter. For this reason, by analyzing $\text{Re}\;\Omega_j$ we gain quantitative information about the Josephson current. Expanding $\text{Re} \;\Omega_J$ using Eq.~\eqref{eq:ImRootApprox} for small $t_0 /\left(\epsilon+ \Sigma(0)\right)$ we get
\begin{equation}
\text{Re} \;\Omega_J \approx -  \frac{2t_0^2}{\epsilon+ \Sigma(0)}.
\end{equation}
 As we see from this relation and from Fig.~\ref{fig:characteristicFrequency},  $\text{Re} \;\Omega_J$ depends sensitively  on the gate potential $\epsilon$. By tuning it  close or far from the transition $\epsilon_{crit}$ we have thus a large or small Josephson current. This thus provides  the possibility to control the current via the gate potential. This property enables to use this system as a transistor. Moreover,  
 one can additionally control the direction of the current.
 It depends on the sign of $\text{Re} \;\Omega_J$ and consequently on the sign of $\epsilon-\epsilon_{crit}$ as can be also seen in  Fig.~\ref{fig:characteristicFrequency}.
 
 It is worth to mention that  a finite imaginary part of $\Omega_J$ induces an additional  phase shift  to $\overline I_{R,J }(t)$. For the chosen parameters in Fig.~\ref{fig:characteristicFrequency}, this is rather small as the ratio of imaginary and real part of $\Omega_J$ is of the order of $10^{-2}$. 
Motivated by Eq.~\eqref{eq:Josephson2mode}, we also investigate the relation between the correlation function  $\mathcal C_{RL}(t)= \left<\myC_{0,R}^\dagger \myC_{0,L}\right>_t$ and the Josephson current. This somehow technical analysis is included into appendix~\ref{sec:JospehsonAndCorr}.

We also find an approximate expression describing the main contribution of the time-averaged dot occupation  $\overline N_d(t)$. 
 We find that the main contribution for long times is given by
\begin{equation}
  \overline N_d(t) \approx  \frac{\left| \Omega_J \right|^2}{4 t_0^2}  e^{2 t \text{Im}\;\Omega_j } \left[ n_R + n_L + 2  \cos \Delta\phi  \sqrt{n_L n_R}  \right] .
  \label{eq:dotOccupation}
\end{equation}
This expression also agrees with the exact calculation depicted in Fig.~\ref{fig:overview}(d) in regime I and resembles a main contribution of the moving time-average in regime II. Interestingly, in regime II
the time-averaged dot occupation vanishes for $t\rightarrow \infty$ as a result of  the finite real part of $z_2$. However, there are additional  contributions to $\overline N_d(t)$ so that the dot occupation does not fully vanish.

\section{Conclusions}

The methods which we presented in this article provide an efficient and accurate tool to determine numerically and analytically the coherent dynamics of  Bose-Einstein condensates in the Fano-Anderson model. 
We showed that the Josephson current sensitively depends on the gate potential like in a transistor. 
In particular, we predict a crossover from a regime with a constant dot occupation  to a regime with an oscillating one. This transition is also visible in the Josephson current  between the reservoirs.
The regimes appear as the energy of the reservoir modes is bounded at  $\omega=0$. As a consequence the energy of the eigenstates  generating the dynamics can be complex-valued, which causes a qualitatively different damping depending on the regime.
 Furthermore, we provide analytical expressions for  observables.

Additionally, we demonstrated how to derive an effective non-hermitian Hamiltonian that exactly describes the time evolution in the long-time limit. Its complex eigenvalues as a function of the gate potential become nearly degenerate  at a critical value $\epsilon=\epsilon_c$. This is analogous to the branching behavior in exceptional points and explains the transition between the two regimes of the time evolution in our system.  

An important point to address in the future is how interactions between the particles influence the dynamics. The interactions could be introduced as in Refs.~\cite{Lee2007,Lee2010}, which investigates the equilibrium properties of a bosonic single-impurity Anderson model. Such kind of investigations could reveal the stability of the Josephson current  in the presence of interactions.

\section{Acknowledgments}

The authors gratefully acknowledge financial support
from the DFG Grants  No. BR 1528/8,
No. BR 1528/9, No. SFB 910, No. GRK 1558 and SCHA 1646/3-1.

 \bibliography{transport}

\appendix

\section{Details of the finite-size simulation}

\label{sec:AppFiniteSizeSimulation}

In the CL, the tunneling elements $t_{k,\alpha}$ and the reservoir frequencies $\omega_{k,\alpha}$ are described simultaneously by the tunnel rates $\Gamma_\alpha(\omega)$. For the finite-size simulation we have to separate them again to define the tunneling elements $t_{k,\alpha}$.
To this end,  we split the tunnel rates into
 \begin{equation}
   \Gamma_\alpha(\omega) = \Lambda_\alpha(\omega) \nu_\alpha(\omega),
 \end{equation}
 where $\nu_\alpha(\omega)\equiv \rho_{\alpha,0} \rho(\omega)$ denotes the density of states in the reservoir and $\Lambda_\alpha(\omega)= \lambda_{\alpha,0} \lambda(\omega)$ describes the coupling
 of the  dot and the reservoir modes. In the CL, the density of states diverges, which we achieve formally by $\rho_{\alpha,0} \rightarrow \infty$, while $\rho(\omega)$
 stays constant. Meanwhile, $\lambda_{\alpha,0} \rightarrow 0$ so that we obtain a finite $\gamma_\alpha= \rho_{\alpha,0} \lambda_{\alpha,0}$. In the following, we choose $ \lambda(\omega)= (\omega/\omega_c)^\eta $, $\nu(\omega)= e^{-\omega/ \omega_c}$, and $\rho_{\alpha,0}=k_{max}/\omega_c$. 
 
 The frequencies of the reservoir modes are  taken as $ \omega_{k,\alpha} = - \omega_c \log \frac {k_{max}-k}{k_{max}} $. In doing so, we make sure that the reconstructed density of states $\nu_{r,\alpha}(\omega)= \sum_{k=0}^{k_{max}} \delta(\omega-  \omega_{k,\alpha})$ fullfills
\begin{equation}
 \int_0^{ \omega_{k,\alpha}}   \nu_{r,\alpha} (\omega) d\omega = \int_0^{\omega_{k,\alpha} }  \nu_\alpha(\omega) d\omega.
\end{equation}
 The tunnel elements $t_{k,\alpha}$ are given by
 \begin{equation}
   t_{k,\alpha}^2 = \Lambda_\alpha( \omega_{k,\alpha}) ,
 \end{equation}
 for $k>0$ .
The coupling $t_{0,\alpha}$ is choosen such that
 \begin{equation}
    t_{0 ,\alpha}^2 \equiv t_{0 }^2  \equiv   \Lambda_\alpha\left(\frac 1 {\rho_0}\right) ,
    \label{eq:GScoupling}
 \end{equation}
 as $ \Lambda_\alpha\left(\omega_{0\alpha}=0\right)=0$ for $\eta>0$. In doing so, we make sure that the ground state is coupled in the same manner as the excited states close to it.
 Consequently,  for an increasing density of states $\rho_{\alpha,0}$, the coupling between dot and reservoir ground states decreases.

\section{Rotation of the branch cut }
 \label{sec:AppendixBranchCutRotation}

Here we show, that the modified function $\tilde C_\alpha(z)$ in Eq.~\eqref{eq:branchCutRotated} is analytic on the negative imaginary axis. For a notational reason we define  $R =  \left\lbrace z \in \mathbb C|\text{Re}\;z =0 \land \text{Im}\; z < 0 \right\rbrace$ which is the negative imaginary axis,
and  $G= \mathbbm C \setminus \left\lbrace z \in\mathbbm C|\text{Re}\;z <0 \land \text{Im}\; z = 0 \right\rbrace $ which is the domain of $\tilde C_{\alpha}$.
We assume that  $C_\alpha(z)$ is analytic in the regions $A_1$ and $A_2$ defined by
\begin{align}
	A_1 &= \left\lbrace z \in G|\text{Re}\;z < 0 \land \text{Im}\; z <0 \right\rbrace, \\
	A_2 &= G \setminus (A_1 \cup R) .
\end{align}
 Furthermore, we assume, that the analytic continuation of 
\begin{equation}
	 \Gamma_\alpha (i z ) \equiv \lim_{\delta \rightarrow 0} \left[ C_\alpha(z +  \delta )-  C_\alpha(z -  \delta )  \right]
\end{equation}
with $\text{Re} \;\delta>0$
is analytic for $z\in A_1 \cup R$. Consequently,  $\tilde C_\alpha(z)$ as defined in \eqref{eq:branchCutRotated} is analytic on $ A_1 $  as it is a sum of analytic functions.
Additionally, we assume that for all derivatives 
\begin{equation}
	  C_\alpha ^{(n)}(z) \equiv \frac{d^n}{dz^n}   C_\alpha ^{(n)}(z) 
 \end{equation}
with $n \in \mathbb N$ the  limit
\begin{equation}
	 f^{(n)}_{j}(z ) = \lim_{\delta \rightarrow 0}  C_\alpha^{(n)}(z +(-1)^j \delta )
\end{equation} 
 with $j=1,2$ exists
 for all $z\in R$. Under these requirements,  we can now show that $\tilde C_\alpha(z)$ is indeed analytic for $z \in R$.

To this end, we show that all derivatives $\tilde C_\alpha ^{(n)}(z)$
are continuous for $z \in R$. Therefore,  we consider for $z=i\omega \in R$ the limit
\begin{align}
\lim_{\delta\rightarrow 0} \tilde C_\alpha ^{(n)}&(-i \omega -\delta) \nonumber \\
&=  \lim_{\delta \rightarrow 0}\left[ C_\alpha ^{(n)}(-i \omega -\delta)  + \Gamma_\alpha^{(n)} (\omega - i \delta) \right] \nonumber  \\
&=   f^{(n)}_{1}(-i\omega ) +   \Gamma_\alpha^{(n)} (\omega )  ,
\label{eq:limitLeft}
\end{align}
where we have used that $\Gamma_\alpha(i z) $ is analytic for $z\in R$.
We continue to calculate 
\begin{align}
 \Gamma_\alpha^{(n)} (\omega ) &\equiv \frac{d^n}{dz^n} \Gamma_\alpha^{(n)} (i z ) \mid_{z=-i\omega} \nonumber \\
 &= \frac{d^n}{d(- i \omega )^n} \lim_{\delta \rightarrow 0} \left[ C_\alpha(- i \omega +  \delta )-  C_\alpha(- i \omega -  \delta )  \right]\nonumber \\
 &= \lim_{\delta \rightarrow 0} \left[ C_\alpha^{(n)}(- i \omega +  \delta )-  C_\alpha^{(n)}(- i \omega -  \delta )  \right] \nonumber \\
 &= f^{(n)}_{2}(-i\omega )-f^{(n)}_{1}(-i\omega ).
\end{align}
Inserting this into Eq.~\eqref{eq:limitLeft} we find
\begin{align}
	\lim_{\delta\rightarrow 0} \tilde C_\alpha ^{(n)}(-i \omega -\delta)& = f^{(n)}_{2}(-i\omega ) \nonumber \\
	 & =\lim_{\delta\rightarrow 0} \tilde C_\alpha ^{(n)}(-i \omega +\delta),
\end{align}
which proves that $ \tilde C_\alpha(z)$ is analytic on the negative real axis and consequently also analytic on $G$.

 \section{Estimation of the branch-cut integral}
 
 \label{sec:AppBranchCutApproximation}
 
 Here we derive an estimate for the branch-cut integral corresponding to the term with $k,k'=0$ in the second line of Eq.~\eqref{eq:SolutionLaplaceC}. The branch-cut integral $\mathcal I_{bc}$ reads
 \begin{align}
	 \mathcal I_{bc} =&\frac{c_{0,\alpha'}}{2\pi i} \int_{-\infty}^{0} \frac {t_{0,\alpha}t_{0,\alpha'} \; e^{x t}  }{(x+ i \omega_{0,\alpha})(x+ i \omega_{0,\alpha'})} \nonumber \\
	  &\times\left[ \frac 1{ \mathcal{\tilde F}(x+ i 0^+)  }  - \frac 1{ \mathcal{\tilde F}(x- i 0^+)  }  \right] dx.
	 \label{eq:branchCutintegral}
 \end{align}
   where
   \begin{equation}
   \mathcal{\tilde F}(x\pm i 0^+) = x+ i \epsilon + \sum_{\alpha} \frac{t_{0,\alpha}^2}{ x+ i \omega_{0,\alpha}}     +  \tilde C_\alpha(x\pm i 0^+).
   \end{equation}
   As $t\rightarrow \infty$, the integrand vanishes everywhere in the long-time limit except at $x= 0$. Therefore, we investigate the integrand close to that point in the following.
   
   From Eq.~\eqref{eq:branchCutRotated} we see that
   \begin{equation}
	   C_+(x) \equiv\sum_{\alpha } \tilde C_\alpha(x+ i 0^+) = \sum_{\alpha } \tilde C_\alpha(x- i 0^+)- \Gamma_\alpha(ix).
   \end{equation}
   Inserting this into \eqref{eq:branchCutintegral}, we find
   \begin{align}
	   \mathcal I_{bc} = \frac{c_{0,\alpha'}}{2\pi i}\int_{-\infty}^{0} &\frac {t_{0,\alpha}t_{0,\alpha'}  \sum_{\alpha} \Gamma_\alpha(ix) }{(x+ i \omega_{0,\alpha})(x+ i \omega_{0,\alpha'})}  \nonumber \\
	    &\times \frac{e^{x t} }{\mathcal{\tilde F}(x+ i 0^+) \mathcal{\tilde F}(x- i 0^+) } dx.
	   \label{eq:branchCutSimplified}
   \end{align}
   To find the leading contribution of $\mathcal I_{bc}$, we have to  approximate the terms
   \begin{equation}
   (x+ i \omega_{0,\alpha'}) \mathcal{\tilde F}(x\pm i 0^+)
   \end{equation}
   appearing in the nominator.
   
   To this end, we expand Eq.~\eqref{eq:genericContinuum}
   for $z\approx0$.
 First we expand the incomplete Gamma function 
 \begin{equation}
 	\tilde \Gamma(-\eta,z)=  \tilde \Gamma(-\eta)+ \frac{z^{-\eta}}{\eta}+ z^{-\eta}\mathcal O(z).
 \end{equation}
 Inserting this into Eq.~\eqref{eq:genericContinuum}  we  obtain
 \begin{align}
 	C_+(z)=&-i \frac{\gamma}{\pi} (- i z/\omega_c )^\eta \tilde \Gamma(1+\eta)\cdot \tilde \Gamma(-\eta) \nonumber \\
 	      & -i \frac{\gamma}{\pi} \Gamma( \eta)   
 	      -i \frac{\gamma}{\pi}  \tilde \Gamma(1+\eta) \mathcal O(z).
 	      \label{eq:CExpansion}
 \end{align}
 For $\eta>0$, the second term dominates, which can be identified in this case with  the Lamb shift in Eq.~\eqref{eq:LampShift}. For $\eta<0$ the function $C_+(z)$ diverges at $z=0$ due to the first term.

\subsection{Case: $\omega_{0,\alpha}=\omega_{0,\alpha'}=0 $} 

In this case,  we find for small $x$
 \begin{equation}
	 x \mathcal{\tilde F}(x\pm i 0^+) \approx  \sum_{\alpha=L,R} t_{0,\alpha}^2=2 t_0^2,
 \end{equation} 
 as we have $\eta>-1$. Here and in the following we assume symmetric tunneling elements $t_{0,L}=t_{0,R}\equiv t_0$ for simplicity. We recall that for $\eta<-1$ the integral~\eqref{eq:continuum} diverges for $z=0$. Inserting this into \eqref{eq:branchCutSimplified} and using that $\Gamma_\alpha(\omega)\approx\gamma_\alpha (\omega/\omega_c)^\eta$, we obtain
    \begin{equation}
 	   \mathcal I_{bc} \approx \frac{c_{0,\alpha'}}{2\pi i}\int_{-\infty}^{0} \frac {   \sum_{\alpha}\gamma_\alpha  }{4 t_0^2}\left(\frac{-ix}{\omega_c}\right)^\eta  e^{x t} dx.
 	   \label{eq:branchCutAppr}
    \end{equation}
 This integral can be analytically solved. In doing so, we get
      \begin{equation}
   	   \mathcal I_{bc} \approx \frac{c_{0,\alpha'}}{2\pi i}  \frac {  \sum_{\alpha}\gamma_\alpha    }{4 t_0^2  }\left(\frac{-i}{\omega_c}\right)^\eta \tilde \Gamma(\eta+1)  \frac 1{ t^{\eta+1}  },
   	   \label{eq:branchCutAppr2}
      \end{equation}
   which is expression~\eqref{eq:branchCutAppr0}.
   
 \subsection{Case: $\omega_{0,\alpha'}\neq \omega_{0,\alpha}=0 $} 
 
 In this case, we find
    \begin{align}
    (x+ i \omega_{0,\alpha'}) \mathcal{\tilde F}(x\pm i 0^+) &\approx  \frac{t_0^2   i \omega_{0,\alpha'} }{ x }  , \\
     x \mathcal{\tilde F}(x\pm i 0^+) &\approx  t_0^2,
    \end{align}
 where we have again used that $\eta>-1$. Inserting this into  \eqref{eq:branchCutSimplified} we get
 \begin{align}
 	   \mathcal I_{bc} &\approx\frac{c_{0,\alpha'}}{2\pi i} \int_{-\infty}^{0} \frac {   \sum_{\alpha} \gamma_\alpha  }{  t_0^2 \omega_{0,\alpha'}}\left(\frac{-ix}{\omega_c}\right)^{\eta}(-ix) e^{x t} dx  \nonumber \\
 	   &=\frac{c_{0,\alpha'}}{2\pi i} \frac { -i \sum_{\alpha} \gamma_\alpha    }{ t_0^2 \omega_{0,\alpha'} }  \left(\frac{-i}{\omega_c}\right)^\eta \tilde \Gamma(\eta+2)  \frac 1{ t^{\eta+2}}.
 	   \label{eq:branchCutAppr3}
 \end{align}

  \subsection{Case: $\omega_{0,\alpha'}\neq 0 $ and $ \omega_{0,\alpha}\neq 0 $} 
  
  Here we find
      \begin{equation}
      (x+ i \omega_{0,\alpha'}) \mathcal{\tilde F}(x\pm i 0^+) \approx 
      \begin{cases}
	      K & \eta>0\\
	      K' x^\eta & \eta<0
      \end{cases} ,
      \end{equation}
   where $K, K'$ are constants and depend on the system parameters. Inserting this into \eqref{eq:branchCutSimplified} we  get
      \begin{equation}
   	   \mathcal I_{bc} \approx  \frac{c_{0,\alpha'}}{2\pi i}  \frac {   \sum_{\alpha} \gamma_\alpha    }{K  }\left(\frac{-i}{\omega_c}\right)^\eta \tilde \Gamma(\eta+1)  \frac 1{ t^{\eta+1}  }
   	   \label{eq:branchCutAppr4}
      \end{equation}
  for $\eta>0$ and 
       \begin{equation}
    	   \mathcal I_{bc} \approx  \frac{c_{0,\alpha'}}{2\pi i}  \frac {   \sum_{\alpha} \gamma_\alpha     }{K'  } \left(\frac{-i}{\omega_c}\right)^\eta \tilde \Gamma(\left| \eta\right| +1)  \frac 1{ t^{\left|\eta\right|+1}  }
    	   \label{eq:branchCutAppr5}
       \end{equation}
   for $\eta<0$.

In a similar way, one can show that all other branch-cut integrals in Eq.~\eqref{eq:SolutionLaplaceD} and  Eq.~\eqref{eq:SolutionLaplaceC} vanish even faster as a function of $t$.

 \section{Derivation of the roots}
 
 \label{sec:AppendixRootsDerivation}
 
 In the following we derive the approximate expression for the location of the roots of $\mathcal{\tilde F}(z)$ in Eq.~\eqref{eq:rootApprox}.
 The procedure is performed in two steps. In the first one, we determine the leading order of the imaginary parts which can be used to determine subsequently in the second step the leading order of the real part.
 
 The root which is  located at $z=0$ for $t_{0,\alpha}\equiv t_0=0$ and $\omega_{0,R}=\omega_{0,L}=0$ is only slightly shifted for a small but finite $t_0$. For this reason, we evaluate $\tilde C_\alpha(z)$ at $z=0$. Assuming additionally $\eta>0$, we thus obtain from  
 \begin{equation}
 0=\mathcal{\tilde F}(z)=   z + i \epsilon +  \frac{2 t_{0 \alpha}^2}{z}  +  \sum_{\alpha =L,R}\tilde C_\alpha (z)  
 \end{equation}
   the quadratic equation
 \begin{equation}
 (z+i\epsilon)z + 2 t_0^2 + z i\Sigma(0)=0 ,
 \label{eq:QuadraticEq}
 \end{equation}
  with $\Sigma(0)$ defined by \eqref{eq:DampingAndLampshift}. Note that $\Gamma(0)=0$ for $\eta>0$.
 This equations has the roots
 \begin{align}
 	 z_{1,2}^{0} &= - i\frac 1 {2 } \left(\epsilon + \Sigma(0) \pm \sqrt{\left( \epsilon + \Sigma(0)\right)^2 + 8 t_0^2 }\right) .
 \end{align}
 We remark that by setting  $\Sigma(0)=0$ we obtain the energies of the  three-mode system without coupling to the excited reservoir modes. 
 We emphasize that the $z_j^{(0)}$ are purely imaginary and thus the leading order of the imaginary part of $z_j$.
  
  To determine the leading order of the real parts, we have to determine the next order  $z_j^{(1)}$ of the roots. To this end, we define
  \begin{equation}
  	\sum_{\alpha =L,R}\tilde C_\alpha (z) \equiv i \Sigma(0) + \tilde C_r(z).
  \end{equation}
  Inserting  $z_{j}=z_j^{(0)}+ z_j^{(1)}$ into $\mathcal{\tilde F}(z)=0$ we get
  \begin{align}
  	(z_j^{(0)}+ z_j^{(1)} &+i\epsilon)(z_j^{(0)}+ z_j^{(1)}) + 2 t_0^{(2)} + \\
  	& (z_j^{(0)}+ z_j^{(1)})\left[ i\Sigma(0)+\tilde  C_r(z_j^{(0)}+ z_j^{(1)})  \right]=0  ,\nonumber
  \end{align}
  which is  an exact relation. As before we approximate the argument of $\tilde C_r(z_j^0+ z_j^1)  \rightarrow\tilde  C_r(z_j^0) $ which is assumed to be small.
  Furthermore we omit the terms $\left(z_j^1\right)^2$ and $\tilde C_r(z_j^0)  z_j^1$ and arrive at a linear equation with the solution
  \begin{equation}
  z_j^1 = \frac {z_j^0  \tilde C_r(z_j^0) }{  2z_j^0 + \left[i \epsilon + \Sigma(0)\right] },
  \end{equation}
  We are interested in its  real part as we have already identified the leading order of the imaginary part $z_j^{(0)}$. 
 Using thus  Eq.~\eqref{eq:DampingAndLampshift}, we finally obtain Eq.~\eqref{eq:rootApprox}.

 \section{Derivation of the time-averaged current}
 
 \label{sec:AppendixTimeAveragedCurrent}

 The exact expression for the current reads
  \begin{align}
  	 I_{R,J} (t)    &=  -2 \text{Re} \; i t_{0,R}\left<{\mathbf v}_{t,3} ^\dagger   {\mathbf v}_{t,1}  \right>_0 \nonumber \\
  		&=  2 \text{Re} \sum_{j,j'}e^{(z_j^*+z_{j'})t}  \tilde  I_{R,J}^C  (z_j,z_{j'}) \label{eq:CurrentExact},\\
  	\tilde I_{R,J}^C  (z_j,z_{j'})&\equiv -i t_{0,R}   \sum_{l,l'}Q_{3,l}^*(z_j) Q_{1,l'}(z_{j'})    \left<{\mathbf v}_{0,l}^\dagger   {\mathbf v}_{0,l'}  \right>_0 ,
  	\label{eq:CurrentFourier2}
  \end{align}
   where $ {\mathbf v}_{t,j}$ is define in Eq.~\eqref{eq:DefV}.
   We recall that one root $z_3$ of $\tilde {\mathcal F}(z) $ converges for the symmetric system $\omega_{0,R}\rightarrow\omega_{0,L}=0$ to $z_3=0$.
   The oscillations are generated by the exponential factor $e^{(z_j^*+z_{j'})t}$. We found that $\tilde  I_{R,J}^C  (z_j,z_{j})=0$.
  The largest period is given by the imaginary part of $z_3^*+z_1$ in regime I and by the imaginary part of $z_3^*+z_2$ in regime II as can be seen in Fig.~\ref{fig:roots}(e). The other root differences are orders of magnitude larger, away from the transition at  $\epsilon = \epsilon_{crit}$.
  For a notational reason we thus introduce in Eq.~\eqref{eq:CharacteristicFrequecy} the characteristic frequency $\Omega_J$.
  
 We define the time-averaged current $\overline I_{R,J }(t)$   by neglecting all other frequency contributions in Eqs.~\eqref{eq:CurrentExact}.
 In doing so, the complex current in both regimes reads
 \begin{align}
 \overline I_{R,J}&(t)=\label{eq:currentApprox}\\ &-2 \text{Re} \; i t_{0}  e^{-i \Omega_J t}   \sum_{l,l'}Q_{3,l}^*(0) Q_{1,l'}(-i \Omega_J) 
     \left<{\mathbf v}_{0,l}^\dagger  {\mathbf v}_{0,l'}  \right>_0  . \nonumber 
 \end{align}
 To show this, one also has   to take into account that 

 \begin{align}
 	&Q( z_3\rightarrow 0) = 
 	\left(
 		\begin{array}{ccc}
 				0                                   & 0  & 0\\
 		 0& \frac 12 & -\frac 12 \\
 	    0 & -\frac 12 &\frac 12
 		\end{array}
 	\right).
 	\label{eq:QMatrixz0}
 \end{align}
 The matrix elements of  $Q_{1,2}(-i \Omega_J)= Q_{1,3}(-i \Omega_J)$ defined in Eq.~\eqref{eq:QMatrix} can be approximated by
 \begin{align}
 	 Q_{1,2}(-i \Omega_J) &= -\frac{ t_0 }{ \Omega_J } R_{-i \Omega_J}  =  \frac{ t_0 }{\Omega_J } \frac1{\left.\frac d{dz} \mathcal{ \tilde F}(z)\right|_{z=-i \Omega_J}} \nonumber \\
 	 &= - \frac{ t_0 }{\Omega_J } \frac 1 { 1+  \frac{ 2 t_0^2}{\Omega_J^2}   +  \frac d{dz}  \sum_{\alpha } \left.\tilde C_\alpha(z)\right|_{z=-i \Omega_J} } \nonumber \\
 	 &\approx - \frac{\Omega_J}{2 t_0}.
 \end{align}
 The approximation  is justified as $\Omega_J$ is small so that $2 t_0^2 / \Omega_J^2$  is large compared to the other terms in the nominator. Inserting this into \eqref{eq:currentApprox} and using Eqs.~\eqref{eq:InitalCorrelation}, we finally obtain Eq.~\eqref{eq:CurrentApprox}.
In a similar manner, we also derive the time-averaged dot occupation in Eq.~\eqref{eq:dotOccupation}.

 \section{Josephson current and correlation function}
 \label{sec:JospehsonAndCorr}
 
 In the following, we establish a relation between the Josephson current $I_{R,J}(t)$ and the correlation function 
 $$\mathcal C_{RL}(t)\equiv \left<\myC_{0,R}^\dagger \myC_{0,L}\right>_t  , $$
  in order to generalize Eq.~\eqref{eq:Josephson2mode}. However, the relation can not be expressed in a simple way in the time domain as in~\eqref{eq:Josephson2mode}, but has to be done in Fourier space.
  
  Motivated by the theoretical description of electronic systems, we define the complex current operator
  \begin{equation}
  \mathbf{ I}_{\alpha,J}^C =  - i  t_{0,\alpha} \myC_{0,\alpha}^\dagger \myD .
  \end{equation}
 The physical current $I_{R,J}$ is given by  $I_{R,J} (t)=2\;\text{Re}\; \left<\mathbf{ I}_{\alpha,J}^C \right>_t $.
 Using \eqref{eq:CurrentExact}, we find that the complex current reads
 \begin{align}
 	 I_{R,J}^C (t)    &=  -i t_{0,R}  \left<{\mathbf v}_{t,3} ^\dagger   {\mathbf v}_{t,1}  \right>_0 \nonumber \\
 		&= \sum_{j,j'}e^{(z_j^*+z_{j'})t}  \tilde  I_{R,J}^C  (z_j,z_{j'}) .
 	\label{eq:CurrentFourier}
 \end{align}
 The constants $	\tilde I_{R,J}^C  (z_j,z_{j'})$ are the Fourier components and are given in Eq.~\eqref{eq:CurrentFourier2}. To link the current Fourier components to  the correlation function $\mathcal C_{RL}(t)$,
 we express it using its Fourier components
 \begin{align}
 	 \mathcal C_{RL} (t)  &=   \left<{\mathbf v}_{t,3} ^\dagger   {\mathbf v}_{t,2}  \right>_0 \nonumber \\
 		&= \sum_{j,j'}e^{(z_j^*+z_{j'})t}  \tilde{ \mathcal C}_{RL}  (z_j,z_{j'}), \\
 	 \tilde{ \mathcal C}_{RL}  (z_j,z_{j'})&\equiv \sum_{l,l'} Q_{3,l}^*(z_j) Q_{2,l'}(z_{j'})    \left<{\mathbf v}_{0,l}^\dagger   {\mathbf v}_{0,l'}  \right>_0 .
 	\label{eq:CorrelationFourier}
 \end{align}
  Now we recognize that
 \begin{equation}
   Q_{1,l'}(z_{j'})  =-\frac 1{t_{0,L}} \left(\omega_{0,L} -i z_{j'} \right)  Q_{2,l'}(z_{j'}),
 \end{equation}
 which is obvious from Eq.~\eqref{eq:QMatrix}.  Inserting this relation in Eq.~\eqref{eq:CurrentFourier2}, we finally obtain
 \begin{align}
 	\tilde I_{R,J}^C (z_j,z_{j'}) & = \kappa_j   \mathcal{ \tilde  C}_{RL} (z_j,z_{j'}), \\
 	\kappa_{j'} &=  i \frac{t_{0,R}}{t_{0,L}}\left(\omega_{0,L} -i z_{j'} \right).
 	\label{eq:CurrentCorrExact}
 \end{align}
 This constitutes an exact relation between the current and the correlation function in Fourier space and therefore generalizes Eq.~\eqref{eq:Josephson2mode}.
 While the $ \tilde I_{R,J}^C$ and $\mathcal{ \tilde  C}_{RL} $ depend on the initial state, the proportional factor $\kappa_j$ depends only on the system parameters. Consequently, the factors $\kappa_j$ characterize  the current through the transistor as a response to the correlation function. Expressing the current as a function of time, we finally obtain
    \begin{equation}
    I_{R,J}(t) = 2 \text{Re} \sum_{j,j'}e^{(z_j^*+z_{j'})t} \kappa_{j'}  \mathcal{ \tilde  C}_{RL} (z_j,z_{j'}).
    \end{equation}

\end{document}